\documentstyle[eqsecnum,aps,12pt]{revtex}
\def\beq{\begin{equation}}
\def\eeq{\end{equation}}

\def\bea{\arraycolsep .1em \begin{eqnarray}}
\def\eea{\end{eqnarray}}

\def\vp{{\bf p}}
\def\vx{{\bf x}}

\def\vv{{\bf v}}

\def\Tr{{\rm Tr}}

\let\de=\delta
\let\eps=\epsilon

\def\bpi{\mbox{\boldmath$\pi$}}
\def\bphi{\mbox{\boldmath$\phi$}}

\def\eq#1{(\ref{#1})}

\def\s0#1#2{\mbox{\small{$ \frac{#1}{#2} $}}}
\def\0#1#2{\frac{#1}{#2}}

\def\llangle{\left\langle}
\def\rrangle{\right\rangle}

\begin{document}
\begin{center}

\thispagestyle{empty}

{\normalsize\begin{flushright}
CERN-TH-99-151\\ 
ECM-UB-PF-99-12\\
{\tt hep-ph/9906210}\\[12ex] 
\end{flushright}}

\mbox{\large \bf 
{Effective Transport Equations for non-Abelian Plasmas}
}\\[6ex]

{Daniel F. Litim
\footnote{E-Mail: Litim@ecm.ub.es}${}^{,a}$ 
and Cristina Manuel
\footnote{E-Mail:  Cristina.Manuel@cern.ch}${}^{,b}$}
\\[4ex]
{\it ${}^a$Departament ECM {\rm \&} IFAE,
Facultat de F\'{\i}sica\\ Univ. de Barcelona,
Diagonal 647, E-08028 Barcelona, Spain.\\[2ex]
${}^b$Theory Division, CERN, CH-1211 Geneva 23, Switzerland.}
\\[10ex]
 
{\small \bf Abstract}\\[2ex]
\begin{minipage}{14cm}{\small
Starting from classical transport theory, 
we derive a set of covariant
equations describing the dynamics of mean 
fields and their statistical fluctuations
in a non-Abelian plasma in or 
out of equilibrium. A general procedure is detailed 
for integrating-out the
fluctuations as to obtain the effective transport equations for the 
mean fields. In this manner, collision integrals for 
Boltzmann equations are obtained as correlators of fluctuations. 
The formalism is applied to a hot non-Abelian plasma close to equilibrium. 
We integrate-out explicitly the fluctuations with typical momenta of the 
Debye mass, and obtain the collision integral in a leading 
logarithmic approximation. We also identify a source for stochastic 
noise. The resulting dynamical equations 
are of the Boltzmann-Langevin type. While our approach
is based on classical physics, we also give the necessary 
generalizations to study the quantum plasmas. Ultimately, the dynamical 
equations for soft and ultra-soft fields
change only in the value for the Debye mass.}
\end{minipage}
\end{center}

\newpage
\pagestyle{plain}
\setcounter{page}{1}

\section{Introduction} 

This article  presents in full detail an approach to the dynamics of 
non-Abelian plasmas based on classical transport theory, the main 
results of which have been summarized in \cite{LM}.

In the recent years, there has been an increasing interest in the 
dynamics of non-Abelian plasmas at very high temperatures or high 
densities. Due to the asymptotic freedom of quantum chromodynamics, 
one expects that quarks and gluons are no longer confined under such 
extreme conditions, but rather behave as free entities forming the 
so-called quark-gluon plasma. Within the next few years, a lot of 
efforts will be given to detect experimentally this new state of 
matter using heavy-ion colliders. 
Another domain of application concerns the physics of the early universe. 
If baryogenesis can finally be understood within an electroweak scenario, 
an understanding of the physics of the electroweak model in the high
 temperature regime where the spontaneous broken symmetry is restored, 
is essential for a computation of the rate of baryon number violation.

It is therefore mandatory to devise reliable theoretical tools for a 
quantitative description of non-Abelian plasmas both in or out of 
equilibrium. While some progress has been achieved in the recent 
years \cite{Gross,ElzeHeinz}, we are still far away from having a 
satisfactory understanding of the relevant relaxation and transport 
processes in non-Abelian plasmas, in particular when it comes to 
out-of-equilibrium situations.

There are different approaches in the literature to study non-Abelian 
plasmas, ranging from thermal field theory to quantum transport 
equations or lattice studies. Even in the close-to-equilibrium plasma, 
and for small gauge coupling, the situation is complicated due to the 
non-perturbative character of long-wavelength excitations in the plasma. 
Most attempts to tackle this problem are based on a quantum field 
theoretical description of the non-Abelian interactions \cite{HTL}.
It has been conjectured that 
the plasma close to equilibrium  allows for a description in terms 
of soft classical fields, as the occupation number for the soft 
excitations are large. While a classical transport theory 
approach  for the non-Abelian case \cite{Heinz} is known to reproduce 
the one-loop thermal effective action \cite{KLLM}, it has never been 
exploited in full detail.  The opposite holds true for 
Coulomb plasmas, where all the essential transport 
phenomena have been studied longly using techniques developed 
within (semi-)classical kinetic theory \cite{K}, while a quantum 
field theoretical approach has been undertaken only recently.

Our approach aims at filling this gap in the literature of classical 
non-Abelian plasmas. Here, we follow the philosophy of 
Klimontovitch \cite{K}, and our equations can be seen as the 
generalization of classical kinetic theory for Abelian plasmas 
to non-Abelian ones. Our essential contribution is considering 
the non-Abelian colour charges as dynamical variables and introducing 
the concept of ensemble average to the non-Abelian kinetic equations. 
Equally important is the consistent treatment of the intrinsic 
non-linearities of non-Abelian gauge interactions.
The observation that Klimontovitch's procedure leads to the 
Balescu-Lenard collision integral for Coulomb plasmas has motivated
earlier derivations of similar (semi-)classical kinetic equations for 
non-Abelian plasmas \cite{Selikhov,SG,Markov}. However, these 
implementations are not fully consistent, and have never been worked out 
in all generality.

The starting point for a classical transport theory of non-Abelian 
plasmas is considering an ensemble of classical point particles 
carrying a non-Abelian charge. They interact through self-consistent 
fields, that is, the fields generated by the particles themselves. 
The microscopic dynamics is governed by the classical equations of 
motion given by the Wong equations \cite{Wong}. When the number of 
particles is large, one has to abandon a microscopic description of 
the system in favour of a macroscopical one based on an ensemble 
average of all the microscopic quantities. This leads naturally to 
a description in terms of averaged quantities, and their statistical 
fluctuations. By averaging the microscopic dynamical equations, we 
obtain effective transport equations for mean quantities. These 
contain the collision integrals of the macroscopic Boltzmann equation, 
which appear in this formulation as statistical correlators of 
fluctuating quantities.  By subtracting the exact microscopic 
equations from the mean ones, we obtain the dynamical equations 
for the fluctuations themselves.  In principle, these two set of 
equations should be enough to consider all the transport phenomena 
in the plasma.

This method is then applied in full detail to a thermal non-Abelian 
plasma close to equilibrium, which allows to employ some approximations. 
For a small plasma parameter, the two-particle correlators are 
small and can be neglected. In the case of small fluctuations, 
the dynamical equations simplify considerably. These conditions 
are always met for a small gauge coupling parameter, which, for 
simplicity, will be assumed throughout. Our approximate equations 
are the leading order ones in a consistent expansion in the gauge 
coupling. However, we shall also see that the condition for a kinetic 
description to be valid could also be met for large gauge couplings. 
After taking statistical averages, we are able to explicitly 
integrate-out the fluctuations with momenta about the Debye mass. 
This gives the collision integrals which appear in the transport 
equations for the mean fields.
In addition to the dissipative processes in the plasma described by 
the collision term, we are able to deduce the stochastic source 
which prevents the system from abandoning equilibrium. This is an 
important result, because it allows to prove explicitly that the 
fluctuation-dissipation theorem holds, when switching from a 
microscopic to a macroscopic description of the system. These 
findings for classical plasmas can be generalized to the case of 
quantum ones. The resulting dynamical equation match perfectly the 
effective theory for the ultra-soft modes as
found by other approaches \cite{Bodeker,Arnold,BI-preprint}.

The lesson to be learned is thus two-fold: There exists a fully 
self-contained formalism to study classical non-Abelian plasmas 
in the first place, which opens in particular a door for applications 
to out-of-equilibrium situations. Second, this approach is -technically 
speaking- much easier than approaches based on the full quantum
field theory. Some of 
the intrinsic complications of a quantum field theoretical description 
(like gauge-fixing, ghost degrees of freedom) can be avoided, and in the close 
to equilibrium plasma, the same effective dynamical equations are obtained.

The paper is organized as follows. We begin with a review of the 
microscopic picture, based on the classical equations of motions for 
coloured point particles (sect.~\ref{Micro}). Changing to a macroscopic 
description needs the introduction of a statistical average, which also allows 
the computation of correlators of fluctuations (sect.~\ref{stati-sec}). 
This procedure is then applied to the fields and the distribution 
function as to obtain dynamical equations for their mean values 
and their fluctuations. The dynamical equations are given in its 
most general form. Possible approximation schemes are detailed, and 
the interpretation of statistical correlators in terms of collision 
integrals is given (sect.~\ref{avera-sec}). The consistency of the
procedure with the requirements of gauge invariance is shown for 
the complete set of equations, and some approximations to 
them (sect.~\ref{sec-consistent}). 
In order to apply the formalism to a plasma close to equilibrium, 
we discuss first the relevant physical scales for both classical 
and quantum plasmas (sect.~\ref{scales}). This is followed by a 
fully detailed derivation of the mean field dynamical equation 
for classical plasmas, which includes the integrating-out of the 
fluctuations with momenta about the Debye mass, and the computation 
of the collision integral and the related noise variable in a leading 
logarithmic approximation. A brief discussion of (iterative) solutions to 
the Boltzmann-Langevin equation for different momentum regimes is also 
presented (sect.~\ref{close2eq}). 
We argue that these results can be translated to the case of 
quantum plasmas and detail the necessary changes. Some comments 
on related work are added as well (sect.~\ref{quantum}). 
Finally, we present our conclusions (sect.~\ref{dis}), deferring to 
the appendices some technical details regarding the Darboux 
variables for SU($N$) colour charges (appendix  \ref{Measure}), and the 
derivation of a useful algebraic identity (appendix \ref{identity}).

\section{Microscopic Equations for Non-Abelian Charged Particles} 
\label{Micro}

Let us consider a system of particles carrying a colour charge $Q^a$, 
where the
colour index runs from $a=1$ to $N^2-1$ for a SU($N$) gauge group. 
Within a microscopic description, the trajectories in phase space are
known exactly. 
The trajectories ${\hat x}(\tau), {\hat p}(\tau)$ and ${\hat Q}(\tau)$ 
for every
particle are
solutions of their classical equations of motions, the Wong 
equations \cite{Wong}
\begin{mathletters}\label{Wong}
\bea
m\0{d{\hat x}^\mu}{d\tau}&=&{\hat p}^\mu \ ,\\
m\0{d{\hat p}^\mu}{d\tau}&=&g {\hat Q}^a F_a^{\mu\nu} {\hat p}_\nu \ ,\\
m\0{d{\hat Q}^a}{d\tau}&=&-g f^{abc} {\hat p}^\mu A^{b}_{\mu}{\hat Q}^c \
.\label{dQ}
\eea
\end{mathletters}%
Here, $A_\mu$ denotes the microscopic gauge field. The corresponding
microscopic field strength $F^{a}_{\mu\nu}$ and the energy momentum tensor
of the gauge fields $\Theta^{\mu \nu}$ are given by
\bea
F^{a}_{\mu\nu}[A]&=&\partial_\mu A^{a}_\mu-\partial_\nu A^{a}_\mu+ gf^{abc}A^{b}_\mu
A^{c}_\nu\ ,\label{NA}\\ \label{NA-EM}
\Theta^{\mu \nu}[A] &=&\s014 g^{\mu \nu}F_{\rho \sigma}^a
F^{\rho \sigma}_a +F^{\mu \rho}_{a} F^{a\, \,\nu}_{\rho}
\eea
and $f^{abc}$ are the structure constants of SU($N$). We set $c=k_B=\hbar =1$ 
and work in natural units, unless otherwise indicated.
Note that the non-Abelian charges are also subject to  dynamical evolution. 
Equation (\ref{dQ}) can
be rewritten as $D_\tau Q = 0$, where
$D_\tau = \frac{ d \hat x^\mu}{d \tau} D_\mu$ is the covariant 
derivative along the world line, and 
$D_\mu^{ac}[A]=\partial_\mu \de^{ac} + g f^{abc} A^b_\mu$ the
covariant derivative in the adjoint representation. With $Q_a$ and
$F^a_{\mu\nu}$ transforming in the adjoint representation, the Wong
equations can be shown to be invariant under gauge transformations.
The equation (\ref{dQ}) ensures the conservation under dynamical evolution
of the set of $N-1$ Casimir of the $SU(N)$ group.\footnote{For $SU(2)$,
it is easy to verify explicitly
the conservation of the quadratic Casimir $Q_a Q_a$. For $SU(3)$, 
both the quadratic
and cubic Casimir $d_{abc} Q_a Q_b Q_c$, where $d_{abc}$ are the 
symmetric structure
constants of the group, are conserved under the dynamical evolution.
The last conservation can
be checked using (\ref{dQ}) and a Jacobi-like identity which 
involves the symmetric $d_{abc}$ and
antisymmetric $f_{abc}$ constants.}

The colour current associated to each particle can be constructed once the
solutions of the Wong equations are known. For every single particle it
reads
\beq\label{j-particle}
j^\mu_a(x) = g \int {d \tau} \frac{d {\hat x}^\mu}{d \tau} \ {\hat Q}_a
(\tau)\ \delta^{(4)}(x -{\hat x}(\tau))  \ .
\eeq
Employing the Wong equations \eq{Wong} we find that $j^\mu$ is
covariantly conserved, $D_\mu j^\mu = 0$ \cite{Wong}.
Similarly, the energy momentum tensor associated to a single 
particle is given by \cite{Wong}
\beq\label{em-particle}
t^{\mu \nu}(x) = \int d \tau \,
\frac{ d{\hat x}^\mu}{d \tau}  \, {\hat p}^\nu(\tau)  
\, \delta^{(4)}(x -{\hat x}(\tau)) \ ,
\eeq

It is convenient to describe the ensemble of particles introducing a 
phase space density  which depends on the whole set of 
coordinates $x^\mu, p^\mu$
and $Q_a$.  We define  the function 
\beq
\label{n-def}
n(x, p, Q) = \sum_{i} \int d \tau \,
\de^{(4)}(x-{\hat x}_i(\tau))\ \de^{(4)}(p-{\hat p}_i(\tau))\
\de^{(N^2-1)}(Q-{\hat Q}_i(\tau))\ ,
\eeq
where 
the index $i$ labels the particles. This
distribution function is constructed in such a way that the colour current
\beq\label{current}
J^\mu _a (x) = g \int d^4p \, d^{(N^2-1)} Q \, \frac{p^\mu}{m} Q_a \, 
n(x,p,Q) 
\eeq
coincides with the sum over all currents associated to the individual
particles,  $J^\mu_a=\sum_ij^\mu_a$, and is covariantly conserved, $D_\mu
J^\mu =0$. 
It is convenient to make  the following
changes in the choice of the distribution function. We will define a new
function $f(x,p,Q)$ such that the physical constraints  like the on-mass 
shell condition, positive energy and conservation of the group Casimirs 
are factored out into the phase space measure. 
We introduce the momentum measure\footnote{Note that in \cite{LM}, 
we used a slightly different normalization of the measure. 
Here, the measure has an
additional factor of $(2 \pi)^3$.}
\beq
\label{Pmeasure}
dP = d^4 p \, 2 \theta(p_0) \de(p^2 -m^2) \ ,
\eeq
and the measure for the colour charges 
\beq
dQ = d^3 Q \, c_R \de (Q_a Q_a -q_2) \ ,
\label{col-mes}
\eeq
in the case of SU($2$). For $SU(3)$ the measure is
\beq
dQ = d^8 Q \, c_R \de (Q_a Q_a -q_2) \de(d_{abc} Q^a Q^b Q^c - q_3) \ .
\label{col-mes-3}
\eeq
For $SU(N)$, $N-1$ $\delta$-functions ensuring 
the conservation of the set of $N-1$ Casimirs
have to be introduced into the measure $dQ$. 
We have also introduced the representation-dependent 
normalization constant $c_R$ into the measure $dQ$, 
which is fixed by requiring  $\int dQ =1$ (see appendix \ref{Measure}).
The constant $C_2$ is defined as
\beq
\label{quadraticQ}
 \int dQ Q_a Q_b = C_2 \delta_{a b} \ ,
\eeq
and depends on the group representation of the particles.

With these conventions the colour current \eq{current} reads now
\beq
\label{j-f}
{J}^\mu_a (x) = g \int dP dQ\, p^\mu \,Q_a \, f(x,p,Q) \ , 
\eeq
while the energy momentum tensor associated to the particles obtains as
\beq\label{em-particles}
t^{\mu \nu}(x) = \int dP dQ\ p^\mu p^\nu\ f(x,p,Q) \ .
\eeq

We will now come to the dynamical equation of the microscopic 
distribution functions 
$n(x,p,Q)$ and $f(x,p,Q)$, which will serve as the starting point for the
subsequent formalism. The dynamical equation for $n(x,p,Q)$ 
is the same as for $f(x,p,Q)$. 
This is so because the physical constraints which we have factored out as to 
obtain $f(x,p,Q)$ are not affected by the Wong equations. 
Employing \eq{Wong}, we find
\begin{mathletters}
\label{NA-Micro}
\beq
\label{NA-f}
p^\mu\left(\0{\partial}{\partial x^\mu}
- g f^{abc}A^{b}_\mu Q^c\0{\partial}{\partial Q^a}
-gQ_aF^{a}_{\mu\nu}\0{\partial}{\partial p_\nu}\right) f(x,p,Q)=0 \ , 
\eeq
which can be checked explicitly by direct inspection of \eq{n-def} into
\eq{NA-f} (see appendix A of \cite{KLLM}). 
In a self-consistent picture this equation
is completed  with the Yang-Mills  equation,
\beq
\label{NA-J}
 (D_\mu F^{\mu\nu})_a(x) =J_a^{\nu}(x) \ ,
\eeq
\end{mathletters}%
and the current being given by \eq{j-f}. It is worth noticing that
\eq{NA-Micro} is {\it exact} in the sense that no approximations 
have been made so far.
The effects of collisions are
included inasmuch as the Wong equations do account for them,
although  \eq{NA-f}  looks formally like a 
{\it collisionless} Boltzmann equation. 

For the energy momentum tensor of the gauge fields we find 
\beq
\partial_\mu \Theta^{\mu \nu} = - F^{\nu\mu}_a J_\mu^a \ .
\label{cons-en-mo}
\eeq
On the other hand, using \eq{NA-f} and the definition \eq{em-particles}
we find that
\beq
\partial_\mu t^{\mu \nu} =\  F^{\nu\mu }_a J_\mu^a  \ .
\eeq
which establishes that the total energy momentum tensor is conserved, 
$\partial_\mu (\Theta^{\mu \nu} +t^{\mu \nu}) = 0$.

To finish the review of the microscopic description of the system, let us
recall the gauge symmetry properties of the set of equations
\eq{NA-Micro} (see \cite{KLLM} for more details).
 From the definition of $f(x,p,Q)$ it follows that it
transforms as a scalar under a (finite) gauge transformation,
$f'(x,p,Q')=f(x,p,Q)$. This implies the gauge covariance of \eq{NA-J}
because the current \eq{current} transforms like the vector $Q_a$ in the
adjoint. The non-trivial dependence of $f$ on the non-Abelian colour
charges implies that the partial derivative $\partial_\mu f$ does not yet
transform as a scalar, but rather its covariant derivative $D_\mu f$,
which is given by
\beq
\label{Df}
D_\mu[A] f(x,p,Q)\equiv 
[\partial_\mu-g f^{abc}Q_c A_{\mu,b}{\partial ^Q_a}]f(x,p,Q)\ .
\eeq
Notice that \eq{Df} combines the first two terms of \eq{NA-f}. 
(Here, and in the sequel we use the shorthand notation 
$\partial_\mu\equiv\partial/\partial x^\mu$,
$\partial_\mu^p\equiv\partial/\partial p^\mu$ and 
$\partial^Q_a\equiv\partial/\partial Q^a$.) The
invariance of the third term in \eq{NA-f} follows from the trivial
observation that $Q_aF^a_{\mu\nu}$ is  invariant under gauge
transformations. This establishes the gauge invariance of \eq{NA-f}.

\section{Statistical averages}
\label{stati-sec}

If the system under study contains a large number of particles it is
impossible to follow their individual trajectories. One has then 
to switch to a statistical description of the system. 

In this section, we describe in detail the statistical average 
to be used in the sequel. As we are studying classical point particles 
in phase space, the appropriate statistical average
corresponds to the Gibbs ensemble average for classical systems.
We will review the main features of this procedure, defined in phase space.
Let us remark that this derivation is completely
general, valid for any classical system, and does not require 
equilibrium situations.

We will introduce two basic functions. The first one is the
phase space density function ${\cal N}$ which gives, after integration over a 
phase space volume element, the number of particles contained in that volume. 
In a microscopic description it is just a deterministic quantity, and a
function of the time $t$, the vectors $\vx $ and $\vp$, and the set
of  canonical (Darboux) variables ${\bphi}$ and ${\bpi}$
associated to the colour charges $Q$ (see appendix \ref{Measure}). 
For $SU(N)$, 
there are $N(N-1)/2$ pairs of canonical variables,
which we denote as ${\bphi} = ( \phi_1, \ldots, \phi_{N(N-1)/2})$ and
${\bpi} = (\pi_1, \ldots, \pi_{N(N-1)/2})$. The microscopic
phase space density is given  by
\beq
\label{NA-PD}
 {\cal N} (\vx, \vp, {\bphi}, {\bpi} ) = 
\sum_{i} \de^{(3)}(\vx-{\hat \vx}_i(t))\
\de^{(3)}(\vp-{\hat \vp}_i(t))\ \de({\bphi} -{\hat {\bphi}}_i(t)) \
\de ({\bpi} - {\hat {\bpi}}_i (t)) \ ,
\eeq
where the sum runs over all particles of the system, and 
$(\hat\vx_i,\hat\vp_i,  \hat{\bphi}_i ,\hat{\bpi}_i)$ 
refers to the trajectory of the $i$-th particle
in phase space. Then 
$ {\cal N} d \vx\, d \vp\, d {\bphi} d{\bpi}$ 
gives the number of particles at time $t$ in
an infinitesimal volume element of phase space around the point 
$z=(\vx,\vp,  {\bphi} ,{\bpi})$

Let us now define the distribution function ${\cal F}$ of the 
microstates of a system of $L$ identical classical particles. Due to 
Liouville's theorem, $d{\cal F}/dt=0$. Thus, it can be normalized as 
\beq
\int dz_1 dz_2 \ldots d z_L\, {\cal F}(z_1,z_2, \ldots , z_L, t) = 1 \ .
\eeq 
Here $z_i$ denotes all the phase space variables associated to the
particle $i$. For
simplicity we have  considered that there is only one species of particles
in the system,
although the generalization to several species of particles is rather
straightforward.

The statistical average of any function ${\cal G}$ defined in phase space
is given by
\beq
\langle {\cal G} \rangle = \int dz_1 dz_2 \ldots d z_L \, {\cal G}(z_1,z_2,
\ldots , z_L) \, {\cal F}(z_1,z_2, \ldots , z_L, t) \ .
\eeq
The one-particle distribution function is 
obtained from ${\cal F}$ as
\beq
 f_1 (z_1,t) = V \int dz_2 \ldots d z_L {\cal F}(z_1,z_2, \ldots , z_L, t) \ ,
\eeq
where $V$ denotes the phase space volume.  Correspondingly, the two-particle 
distribution function is
\beq
 f_2 (z_1,z_2,t) = V^2 \int dz_{3} \ldots d z_L 
\, {\cal F}(z_1,z_2, \ldots, z_L, t) \ ,
\eeq
and similarly for the $k$-particle distribution functions.
A complete knowledge of ${\cal F}$ would allow us to obtain all the set of
$( f_1,  f_2, \ldots , f_L)$ 
functions; this is, however, not necessary for our present purposes.

Notice that we have allowed for an explicit dependence on the time $t$ of
the function ${\cal F}$, as
this would be typically the case in out of equilibrium situations. We will
drop this $t$ dependence
from now on to simplify the formulas.

Using the above definition  one can obtain the mean value
of the microscopic phase space density. Microscopically 
${\cal N}(z) = \sum_{i=1}^L \delta (z -{\hat z}_i)$, 
where ${\hat z}_i$  describes the
trajectory in phase space
of the particle $i$.
The statistical average of this function is
\beq
\langle {\cal N}(z) \rangle = \int dz_1 dz_2 \ldots d z_L {\cal F}(z_1,z_2,
\ldots, z_L)
\sum_{i=1}^L \delta (z -{\hat z}_i) = \frac{L}{V}\, f_1 (z) \ .
\eeq 
The second moment $\langle {\cal N} (z) {\cal N} (z') \rangle$ can equally
be computed, and it is not difficult to see that gives
\beq
\langle {\cal N} (z) {\cal N} (z') \rangle = 
\frac{L}{V} \delta (z-z')  f_1(z) +\frac{L (L-1)}{V^2} f_2(z,z')\ .
\eeq

Let us now define a deviation of the phase space density from its mean
value
\beq
\de {\cal N} (z) \equiv {\cal N} (z) - \langle {\cal N}(z) \rangle \ .
\eeq
By definition  $\langle \de {\cal N}(z) \rangle = 0$, although the second
moment of
this statistical fluctuation does not vanish in general, since
\beq
\langle \de {\cal N} (z) \de {\cal N} (z') \rangle = \langle  {\cal N} (z)
{\cal N} (z') \rangle
-  \langle {\cal N}(z) \rangle \langle {\cal N}(z') \rangle \ .
\eeq
If the number of particles is large, $L \gg 1$, we have
\beq
\langle \de {\cal N} (z) \de {\cal N} (z') \rangle =
\left(\frac{L}{V}\right)\delta (z-z') f_1(z) +
\left(\frac{L}{V}\right)^2 g_2(z,z') \ ,
\eeq
where the function $g_2(z,z') = f_2(z,z') - f_1(z) f_1(z')$ measures the
two-particle correlations in the system. For a completely uncorrelated 
system $g_2 = 0$.

Notice that the above statistical averages are well defined in the 
thermodynamic limit, $L, V \rightarrow \infty$ but $L/V$ remaining 
constant. Higher order moments and higher order correlators can as well 
be defined.

We finally point out that the the function 
${\cal N}(\vx, \vp, {\bphi}, {\bpi} )$ agrees with the 
microscopic function $f(x,p,Q)$ introduced earlier, except for a 
representation-dependent normalization constant $c_R$ as introduced 
in  sect.~\ref{Micro}.  We will also swallow the density factors $L/V$ 
into the mean functions $\bar f$. Those small changes in the 
normalizations allow to simplify slightly the notations of the equations.

\section{Averaging the Microscopic Equations}
\label{avera-sec}

\subsection{The mean fields and the fluctuations}

In this section we perform the step from a microscopic to a
macroscopic formulation of the problem. Using the prescription explained
in sect.~\ref{stati-sec}, we take statistical averages of the microscopic
equations \eq{NA-Micro}.
This implies that the distribution 
function $f(x,p,Q)$,
which in the microscopic picture is a deterministic quantity, has now a
probabilistic nature and can 
be considered as a random function, given by its mean value and statistical
(random) fluctuation about it. Let us define the quantities
\begin{mathletters}
\label{NA-delta}
\bea
f(x,p,Q)&=& {\bar f}(x,p,Q) + \de f(x,p,Q) \ ,\\
J^\mu_a(x)&=&\bar J^\mu _a (x)+\de J^\mu _a(x) \ ,\\
A^{a}_\mu(x)&=& {\bar A}^a_\mu(x) + a^a_\mu(x) \ ,\label{NA-delta-a}
\eea
\end{mathletters}%
where the quantities carrying a bar denote the mean values, e.g. ${\bar
f} = \langle f \rangle$, ${\bar J} = \langle J \rangle$ and ${\bar A} =
\langle A \rangle$, while the mean value of the statistical fluctuations
vanish  by definition,  $\langle \de f \rangle =0$, 
$\langle \de J \rangle =0$ and $\langle a \rangle=0$. 
This separation into mean fields and statistical, random 
fluctuations corresponds effectively 
to a split into low frequency (long wavelength) 
modes associated to the mean fields, and high frequency (short wavelength) 
modes associated to the fluctuations.\footnote{In the close-to-equilibrium 
plasma (sect.~\ref{close2eq}), 
we identify the relevant momentum scales explicitly.}
We also split the field strength tensor as
\begin{mathletters}
\label{NA-deltaF}
\bea
F^{a}_{\mu\nu}& =& {\bar F}^a_{\mu\nu} + f^a_{\mu\nu}\ ,\\
{\bar F}^a_{\mu\nu}&=&F^{a}_{\mu\nu}[\bar A]\ ,\\
f^a_{\mu\nu}&=&(\bar D_\mu a_\nu-\bar D_\nu a_\mu)^a+ g f^{abc}a^b_\mu
a^c_\nu\ ,\label{NA-deltaF-c}
\eea
\end{mathletters}%
using also $\bar D_\mu\equiv D_\mu[\bar A]$. The term $f^a_{\mu\nu}$
contains terms linear and quadratic in the fluctuations. Note that the
statistical average of the field strength  $\langle F^a_{\mu\nu}\rangle$
is given by $\langle F^a_{\mu\nu}\rangle=\bar F^a_{\mu\nu}+
g f^{abc}\langle a^b_\mu a^c_\nu\rangle$, due to quadratic terms contained
in $f^a_{\mu\nu}$.

In the same light, we split the energy momentum tensor 
of the gauge fields into 
the part from the mean fields and the fluctuations,  according to
\begin{mathletters}
\bea
\Theta^{\mu \nu} &=&\bar \Theta^{\mu \nu} + \theta^{\mu \nu}\\
\bar \Theta^{\mu \nu} & = & 
\s014 g^{\mu \nu} \bar F^a_{\rho \sigma}  \bar F_a^{\rho \sigma} +
 \bar F^{\mu \rho}_{a}  \bar F^{a\,\,\nu}_{\rho} \ , \\
\theta^{\mu \nu} & = & 
\s012 g^{\mu\nu}\bar F^a_{\rho\sigma}f^{\rho\sigma}_a
+\bar F^{\mu\rho}_af^a_{\rho\nu}
+\bar F^{\nu\rho}_a f^a_{\rho\mu}
+ \s014 g^{\mu \nu}  f_{\rho \sigma, a} f^{\rho \sigma, a} 
+  f^{\mu \rho}_{a}   f^{a\,\,\nu}_{\rho} \ ,
\eea
\end{mathletters}%
The term $\theta^{\mu \nu}$ contains the fluctuations up to quartic order.
Due to the non-linear character of the theory, we find that the ensemble
average of the energy momentum tensor reads
$\langle\Theta^{\mu \nu}\rangle=\bar \Theta^{\mu \nu} 
+\langle\theta^{\mu \nu}\rangle$.

\subsection{Dynamical equations for the mean fields and the fluctuations}

We perform now the step from the microscopic to the macroscopic Boltzmann
equation, taking the statistical average of (\ref{NA-Micro}). This yields
the dynamical equation for the mean values,
\begin{mathletters}\label{NA-Macro}
\begin{equation}
p^\mu\left(\bar D_\mu- gQ_a\bar F^a_{\mu\nu} 
 \partial _p^\nu\right)\bar f=\left\langle\eta\right\rangle
+ \left\langle\xi\right\rangle \ . \label{NA-1}
\end{equation}
We have made use of the covariant derivative of $f$ as introduced in
\eq{Df}. The macroscopic Yang-Mills equations are obtained as
\begin{equation}
\bar D_\mu \bar F^{\mu\nu}  + \left\langle J_{\mbox{\tiny fluc}}^{\nu}
\right\rangle =\bar J^\nu \ .\label{NAJ-1} 
\end{equation}
\end{mathletters}%
In \eq{NA-Macro}, we collected all terms quadratic or cubic in the
fluctuations into the functions $\eta(x,p,Q), \xi(x,p,Q)$ 
and $J_{{\mbox{\tiny fluc}}}(x)$.
They read explicitly
\begin{mathletters}\label{NA-func}
\begin{eqnarray}
\eta(x,p,Q) 
&\equiv & 
gQ_a\,p^\mu\partial_p^\nu f_{\mu\nu}^a\,\delta f(x,p,Q)\ ,\label{NA-eta}
\\
\xi(x,p,Q) 
&\equiv & 
gp^\mu f^{abc}Q^c\left(\partial ^Q_a a_\mu^b\ \delta f(x,p,Q)\ 
+ g a_\mu^a a_\nu^b\partial^\nu_p\bar f(x,p,Q)\right) ,    \label{NA-xi}
\\
J_{\mbox{\tiny fluc}}^{a,\nu}(x)
&\equiv & 
 g \left\{f^{dbc} \bar D^{\mu}_{ad}  a_{b,\mu} a_c^\nu  
 + f^{abc} a_{b,\mu}\, \left( (\bar D^\mu a^\nu-\bar D^\nu a^\mu)_c
    + g f^{cde}a^\mu_d a^\nu_e \right)\right\}(x)\, .   \label{NA-Jfluc}
\end{eqnarray}
\end{mathletters}%
The corresponding equations for the fluctuations are obtained by  
subtracting (\ref{NA-Macro}) from (\ref{NA-Micro}). The result is
\begin{mathletters}\label{NA-fluc}
\begin{eqnarray}
p^\mu\left(\bar D_\mu 
           - gQ_a\bar F^a_{\mu\nu}\partial _p^\nu\right)\delta f
&=&
g Q_a(\bar D_\mu a_\nu-\bar D_\nu a_\mu)^a p^\mu \partial^p_\nu\bar f
\nonumber \\ && 
+ gp^\mu a_{b,\mu} f^{abc} Q_c\partial ^Q_a \bar f 
+\eta  + \xi - \left\langle\eta+\xi\right\rangle\label{NA-2}\\
\left(\bar D^2 a^\mu-\bar D^\mu(\bar D_\nu a^\nu)\right)^a
&+& 2 gf^{abc}
\bar F_b^{\mu\nu}a_{c,\nu}+J_{{\mbox{\tiny fluc}}}^{a,\mu}-\left\langle
J_{{\mbox{\tiny fluc}}}^{a,\mu}\right\rangle
=\delta J^{a,\mu} \ .\label{NAJ-2}
\end{eqnarray}
\end{mathletters}%
The above set of dynamical equations is enough to describe all 
transport phenomena in the plasma.

While the dynamics of the mean fields \eq{NA-Macro} depends on correlators 
quadratic and cubic in the fluctuations, the dynamical equations for the 
fluctuations \eq{NA-fluc} do in addition depend on higher order terms (up
to cubic order) in the fluctuations themselves. 
The dynamical equations for the higher order correlation functions are
contained in \eq{NA-fluc}. 
To see this, consider for example the dynamical 
equation for the correlators $\langle\delta f\, \delta f\rangle$. 
After multiplying \eq{NA-2} with $\delta f$ and taking the statistical 
average, we obtain
\begin{eqnarray}
p^\mu\left(\bar D_\mu - gQ_a\bar F^a_{\mu\nu}\partial _p^\nu\right)
\langle\delta f\, \delta f \rangle
&=&g Q_a p^\mu \partial ^p_\nu\bar f\ 
\llangle (\bar D_\mu a_\nu-\bar D_\nu a_\mu)^a \delta f \rrangle
\nonumber \\ && \label{quad-corr}
+gp^\mu  f^{abc} Q_c\partial ^Q_a \bar f \ \llangle a_{b,\mu}\,
\delta f\rrangle
+\llangle (\eta  + \xi)\, \delta f\rrangle \ .
\end{eqnarray}
In the same way, we find for $\langle\delta f\, \delta f\, \delta f\rangle$ 
the dynamical equation
\begin{eqnarray}
p^\mu\left(\bar D_\mu - gQ_a\bar F^a_{\mu\nu}\partial _p^\nu\right)
\langle\delta f\, \delta f\, \delta f\rangle
&=&g Q_a p^\mu \partial ^p_\nu\bar f\ 
\llangle (\bar D_\mu a_\nu-\bar D_\nu a_\mu)^a \, \delta f\, \delta f\rrangle
-\llangle\eta  + \xi\rrangle\llangle \delta f\, \delta f\rrangle 
\nonumber \\ &&
+gp^\mu  f^{abc} Q_c\partial ^Q_a \bar f \ 
\llangle a_{b,\mu}\, \delta f\, \delta f\rrangle
+\llangle (\eta  + \xi)\, \delta f\, \delta f\rrangle  \ ,\label{cubic-corr}
\end{eqnarray}
and similarly for higher order correlators. 
Typically, the dynamical equations for correlators of $n$ fluctuations 
will couple to correlators ranging from the order $(n-1)$ 
up to order $(n+2)$ in the fluctuations. 
From cubic order onwards, the back-coupling contains terms 
non-linear in the correlation functions.\footnote{The resulting 
hierarchy of dynamical equations for the correlators 
is very similar to the BBGKY hierarchy. While the BBGKY hierarchy links
the dynamical equations for $n$-particle distribution functions 
with each other, here, we rather have a hierarchy 
for the correlator functions.}

The dynamical equation for the energy momentum tensor of the 
gauge fields obtains from the average of (\ref{cons-en-mo}). The corresponding
one for the particles is found after integrating \eq{NA-f} over 
$dPdQ\, p^\mu$. They read
\begin{eqnarray}
\partial_\nu \bar \Theta^{\mu \nu} 
+ \partial_\nu \llangle \theta^{\mu \nu}\rrangle
&=& - \bar F^{\mu \nu}_a \bar J_{\nu a} 
- \langle f^{\mu \nu}_a \delta J_{\nu a}\rangle 
-  \llangle f_a^{\mu\nu} \rrangle  \bar J_\nu^a \ ,
\\
\partial_\nu \bar t^{\mu \nu} 
&=&\ \ \, \bar F^{\mu \nu}_a \bar J_{\nu a} 
+ \langle f^{\mu \nu}_a \delta J_{\nu a}\rangle 
+  \llangle f_a^{\mu\nu} \rrangle  \bar J_\nu^a \ ,
\end{eqnarray}
such that the total energy momentum tensor is conserved.

Finally, the condition for the microscopic current conservation 
translates, after averaging, into two equations, one for the mean 
fields, and another one for the fluctuation fields. From 
$\langle D_\mu J^ \mu\rangle=0$ we obtain
\beq\label{DJ}
  (\bar D_\mu \bar J^\mu)_a
+ g f_{abc}\langle a^b_\mu \de J^{c,\mu}\rangle
= 0 \ .
\eeq
For the fluctuation current, 
we learn from $D_\mu J^ \mu-\langle D_\mu J^ \mu\rangle=0$ that
\beq\label{DdJ}
(\bar D_\mu \de J^\mu)_a 
+ gf_{abc}\left( a^b_\mu \bar J_c^\mu + a^b_\mu \de J_c^\mu- 
\langle a^b_\mu \de J_c^\mu\rangle\right)=0\ .
\eeq
In sect.~\ref{consistentcurrent}, it is shown that \eq{DJ} 
and \eq{DdJ} are consistent with the corresponding equations 
as obtained from the Yang-Mills equations.

\subsection{Second moment approximation and small coupling expansion}

The dynamical equations, as derived and presented here, are exact. 
No approximations have been performed. In order to solve them, it 
will be essential to  apply some approximations, 
or to find a reasonable truncation for the hierarchy of dynamical equations 
for correlator functions. Here, we will indicate two approximation
schemes, the second moment approximation and the small coupling expansion.
Although they have a distinct origin in the first place, we will see
below (sect.~\ref{sec-consistent}) that they are 
intimately linked due to the requirements of gauge invariance.

The so-called {\it second moment approximation}  (sometimes also 
referred to as the {\it polarization approximation} \cite{K}) is
employed in order to simplify the dynamics of the fluctuations. It 
consists in equating
\beq\label{2nd}
                   \eta=\langle\eta \rangle\ ,\qquad
                   \xi =\langle \xi \rangle\ ,\qquad
J_{{\mbox{\tiny fluc}}}=\langle J_{{\mbox{\tiny fluc}}} \rangle\ .
\eeq 
The essence of this approximation is that the dynamical equations 
for the correlators becomes homogeneous. It is easy to see that 
\eq{quad-corr} or \eq{cubic-corr} depend only on quadratic or cubic 
correlators, respectively, once
\eq{2nd} is imposed. This approximation allows to cut the 
infinite hierarchy of equations down to a closed system of differential 
equations for both mean quantities and statistical fluctuations. 
The mean fields couple to quadratic correlators, and all correlators of
degree $n$ couple amongst each others. This turns the dynamical equation 
for the fluctuations \eq{NA-fluc} into a differential equation linear 
in the fluctuations. This approximation is viable if the fluctuations 
and the two-particle correlators are small (see also sect.~\ref{scales}). 

A seemingly different approximations concerns the non-Abelian sector of 
the theory. We shall perform a systematic perturbative expansion in 
powers of the gauge coupling $g$, keeping only the leading order terms.
This can be done, because  the differential operator appearing in the 
Boltzmann equation \eq{NA-f} or \eq{NA-1} admits such an expansion. 
In a small coupling expansion, the term  
$gQ_a\bar F^a_{\mu\nu}\partial _p^\nu$ is suppressed by a power in
$g$ as compared to the leading order term $p^\mu\bar D_\mu$. 
Notice, that expanding the $g$ appearing within 
$p^\mu\bar D_\mu$ of \eq{Df} is not allowed as it will break gauge 
invariance. In this spirit, we expand as well
\bea
\bar f&=&      \bar f^{(0)}
         +g\,  \bar f^{(1)}
         +g^2\,\bar f^{(2)}
         +\ldots           \label{f-g}
\eea
and similarly for $\de f$. This is at the basis for a systematic 
organization of the dynamical equations in powers of $g$. To leading 
order, this concerns in particular the cubic correlators in 
$\langle\eta\rangle$ and $\langle J_{\mbox{\tiny fluc}}\rangle$, which 
are suppressed by a power of $g$ as compared to the quadratic ones.

In principle, after these approximations are done,
it should be possible to express the correlators of fluctuations 
appearing in \eq{NA-Macro} through known functions. This requires finding
a solution of the fluctuation dynamics first.
 
\subsection{Collision integrals}

Let us comment on 
the interpretation of $\langle\eta\rangle$ and $\langle\xi\rangle$ as
collision integrals of the macroscopic Boltzmann equation. 
The functions $\langle\eta\rangle$ and $\langle\xi\rangle$ appear only
after the splitting \eq{NA-delta} has been performed. This introduces
new terms in the corresponding Boltzmann equation \eq{NA-Macro}
for the mean fields,  which are interpreted as effective 
collision integrals for the macroscopic
transport equation. In this formalism, 
the collision integrals appear naturally as correlators of statistical 
fluctuations. The fluctuations in the gauge fields cause random changes 
in the motion of the particles, while random changes in the distribution
function of the particles 
induce changes in the gauge fields. This is having the 
same effect as collisions, and yields a precise recipe 
as to obtain collision integrals from the microscopic theory.

In this light, the second moment approximation \eq{2nd} can be 
interpreted as neglecting the back-coupling of collisions to the 
dynamics of the fluctuations. Also, neglecting cubic correlators appearing 
in $\langle\eta\rangle$ or $\langle J_{\mbox{\tiny fluc}}\rangle$ 
in favour of quadratic ones, to leading order in an
expansion in the gauge coupling and for small fluctuations, 
is interpreted as neglecting 
the three-particle collisions in favour of two-particle 
collisions.

In order to find explicitly the corresponding collision integrals 
for the non-Abelian plasma, one has to solve first the dynamical 
equations for the fluctuations in the background of the mean fields. 
This step is interpreted as integrating-out the fluctuations. In general, 
this is a difficult task, in particular due to the non-linear 
terms present in \eq{NA-fluc}. As argued above, this will only be possible
when some approximations have been performed.

In the Abelian limit, \eq{NA-Macro} and \eq{NA-fluc} reduce to the 
known set of kinetic equations for Abelian plasmas \cite{K}. In this limit, 
only the collision integral $\langle\eta\rangle$ survives. 
Here it is known  that $\langle\eta\rangle$ can be expressed 
explicitly as the Balescu-Lenard collision integral,  after solving 
the dynamical equations for the fluctuations and computing the 
correlators involved \cite{K}. This proofs in a rigorous way the 
correspondence between fluctuations and collisions in an Abelian plasma.

\section{Consistency with gauge symmetry}
\label{sec-consistent}

In this section we shall discuss the consistency of the present approach
with the requirements of the non-Abelian gauge symmetry. This discussion 
will concern the consistency of the general set of equations. The 
question of consistent approximations will be raised as well. In this 
section, we shall for convenience switch
to a matrix notation, using the conventions $A\equiv A^a t_a$, 
$Q\equiv Q^a t_a$ etc., as well as $[t_a,t_b]=f_{abc}t^c$ and 
$\Tr\ t_at_b = -\s012 \de_{ab}$. 

\subsection{Gauge covariance of the macroscopic equations}

As a consequence of the Wong equations being gauge invariant, we already
established in sect.~\ref{Micro} that the microscopic set of equations
\eq{NA-Micro} transforms covariantly under (finite) gauge transformations
\beq
gA'_\mu = U(x)(\partial_\mu + gA_\mu)U^{-1}(x)\ ,\quad 
U(x)    =\exp -g\eps^a(x)t_a \ ,
\eeq
with parameter $\eps^a(x)$.
We also have $f'(x,p,Q')=f(x,p,Q)$, and
\beq
         Q' = U(x)\,      Q         \, U^{-1}(x)\ ,\quad 
\partial_Q' = U^{-1}(x)\, \partial_Q\, U(x)\      ,\quad 
F_{\mu\nu}' = U(x)\,      F_{\mu\nu}\, U^{-1}(x)\ .
\eeq
The question raises as to which extend this symmetry is conserved under
the statistical average, performed when switching to a macroscopic
description. This concerns in particular the subsequent split of the 
gauge field into a mean (or background) field, and a fluctuation field
\beq\label{split-A}
A_\mu=\bar A_\mu+a_\mu\ .
\eeq 
This separation is very similar to what is done in the background
field method (BFM) \cite{Abbott}. Two symmetries are left after the
splitting is performed, the {\it background gauge symmetry},
\beq\label{BGS}
g \bar A'_\mu=U(x)(\partial_\mu + g\bar A_\mu)U^{-1}(x)\ ,\quad
       a_\mu'=U(x)\, a_\mu\, U^{-1}(x)\ ,
\eeq
and the {\it fluctuation gauge symmetry},
\beq\label{QGS}
g \bar A'_\mu =  0\ ,\quad 
      ga_\mu' =  U(x)\left(\partial_\mu 
               + g(\bar A_\mu+a_\mu)\right)U^{-1}(x)\ .
\eeq
Under the background gauge symmetry, the fluctuation field transforms
covariantly (as a vector in the adjoint). In the first step, we will
split \eq{NA-f} according to \eq{split-A}. It follows trivially, that the
resulting equation is invariant under both \eq{BGS} and \eq{QGS}, if both
$\bar f$ and $\delta f$ transform as $f$, that is, as a scalar. 

The next step involves the statistical average. We
require that the statistical average of the fluctuation vanishes, $\langle
a\rangle=0$. This constraint is fully compatible with the background gauge
symmetry, as $\langle a\rangle=0$ is invariant under \eq{BGS}. Any
inhomogeneous transformation law for $a$, and in particular \eq{QGS}, 
can no longer be a symmetry of the macroscopic equations
as the constraint $\langle a\rangle=0$ is not invariant.\footnote{This is
similar to what happens in the BFM, where the fluctuation gauge
symmetry can no longer be seen once the expectation value
of the fluctuation field is set to zero. However, the symmetry 
\eq{QGS} will be observed in both \eq{NA-1}
and \eq{NA-2}, as long as the terms linear in $\langle a\rangle$ are
retained.} 

We rewrite now the transport equations in matrix convention. We  have
\begin{mathletters}\label{M-NA-Macro}
\begin{eqnarray}
p^\mu\left(\bar D_\mu
  +2 g\ \Tr\,\big( Q\bar F_{\mu\nu}\big)\,\partial _p^\nu\right)\bar f
&=& 
  \left\langle\eta\right\rangle
+ \left\langle\xi\right\rangle \ . \label{M-NA-1}
\\
  \left[ \bar D_\mu,  \bar F^{\mu\nu} \right]  
+ \left\langle J_{\mbox{\tiny fluc}}^{\nu}\right\rangle 
&=& 
  \bar J^\nu \ . \label{M-NAJ-1} 
\end{eqnarray}
\end{mathletters}%
with
\begin{mathletters}\label{M-NA-func}
\begin{eqnarray}
\eta(x,p,Q) &=&
   -2 g \ \Tr\, \big(Q\, f_{\mu\nu}\big)\, p^\mu
                             \partial _p^\nu \delta f(x,p,Q)  \ , 
\label{M-NA-eta}\\
\xi(x,p,Q)&=& 
   -2 gp^\mu\ \Tr\, \big([Q,\partial ^Q]\, a_\mu\big)\, \delta f(x,p,Q)\ 
   -2 g^2\ \Tr\,\big([a_\mu, a_\nu]Q\big)\,p^\mu\partial^\nu_p\bar f(x,p,Q),
\label{M-NA-xi}\\
J_{{\mbox{\tiny fluc}}}^{\nu}(x)&=& 
       g\left[\bar D^{\mu},[a_{\mu},a^\nu]\right]  
   +    \left[a_{\mu},[\bar D^\mu, a^\nu]-[\bar D^\nu, a^\mu]\right]
   + g^2\big[a_\mu,[a^\mu,a^\nu]\big] .\label{M-NA-Jfluc}
\end{eqnarray}
\end{mathletters}%
and
\begin{mathletters}\label{M-NA-fluc}
\begin{eqnarray}
p^\mu\left(\bar D_\mu 
           - g\ \Tr\,\big(Q\bar F_{\mu\nu}\big)\,\partial_p^\nu\right)
\delta f&=&-2 g \Tr\,\big(Q[\bar D_\mu, a_\nu]-Q[\bar D_\nu, a_\mu]\big)
\, p^\mu\partial ^p_\nu\bar f 
\nonumber\\ 
&&
-2 g\, p^\mu \ \Tr\,\big([Q,\partial^Q]a_{\mu}\big)\,\bar f
+\eta  + \xi - \left\langle\eta+\xi\right\rangle \ , \label{M-NA-2}\\
\left[\bar D_\nu,[\bar D^\nu,a^\mu]\right]-\left[\bar D^\mu,[\bar
D_\nu,a^\nu]\right]&+&2 g[\bar F^{\mu\nu},a_{\nu}]+J_{{\mbox{\tiny
fluc}}}^{\mu}-\left\langle J_{{\mbox{\tiny fluc}}}^{\mu}\right\rangle
=\delta J^{\mu} \ .\label{M-NAJ-2}
\end{eqnarray}
\end{mathletters}%
It is now straightforward to realize that \eq{M-NA-Macro} -- \eq{M-NA-fluc}
transform covariantly under \eq{BGS}. It suffices to employ the cyclicity of
 the trace, and to note that $a_\mu$ and background covariant derivatives 
of it  transform covariantly. 

\subsection{Consistent current conservation}\label{consistentcurrent}

In \eq{DJ} and \eq{DdJ}, we have given the equations which 
imply the covariant 
current conservation of the mean and the fluctuation current. 
However, this information is contained both in the transport 
and in the Yang-Mills equation. It remains to be shown that these 
equations are self-consistent.

We start with the mean current $\bar J$. Performing $g\int dPdQ\, Q$ of  
the transport equation 
\eq{M-NA-1}, we find
\beq\label{DJ1}
0=  [\bar D_\mu, \bar J^\mu] 
  + g \langle[a_\mu,\de J^\mu] \rangle\ .
\eeq 
This is \eq{DJ}. Here, we used that
\begin{mathletters}\label{someintegrals}
\bea
   \int dP\ \eta(x,p,Q)                         &=&0 \ ,
\\
\int dP\ p^\mu F_{\mu\nu}\partial_p^\nu f(x,p,Q)&=&0 \ ,
\\
   \int dPdQ\, Q\ \xi(x,p,Q)                    &=& - [a_\mu,\de J^\mu] \ ,
\\
  g\int dPdQ\, Q\ p^\mu \bar D_\mu \bar f(x,p,Q)&=&[\bar D_\mu,\bar J^\mu]\ .
\eea
\end{mathletters}%
Taking the background covariant derivative
of \eq{M-NAJ-1}, we find
\beq
0 = [\bar D_\mu, \bar J^\mu] 
   -[\bar D_\mu,\langle J_{{\mbox{\tiny fluc}}}^\mu \rangle]\ ,
\eeq
which has to be consistent with \eq{DJ1}.
Thus, combining these two equations we end up with the consistency condition
\beq\label{consistencyM}
0= [\bar D_\mu,\langle J_{{\mbox{\tiny fluc}}}^\mu \rangle]
  + g \langle[a_\mu,\de J^\mu] \rangle\ .
\eeq
In the appendix \ref{identity}, we established the identity
\beq\label{general-cb}
0=[\bar D_\mu,J_{{\mbox{\tiny fluc}}}^\mu ]+g [a_\mu,\de J^\mu]
+g[a_\mu,\langle J_{{\mbox{\tiny fluc}}}^\mu \rangle]\ ,
\eeq
which follows using the explicit expressions for
$\de J^\mu$ from \eq{M-NAJ-2} and for $J_{{\mbox{\tiny fluc}}}$ from
\eq{M-NA-Jfluc}.
Taking the average of \eq{general-cb} reduces it to
\eq{consistencyM}, and establishes the self-consistent conservation 
of the mean
current.

The analogous consistency equation for the fluctuation current obtains
from \eq{M-NA-2} after performing $g\int dPdQ\, Q$, and reads
\beq\label{consistency2}
 0=  [\bar D_\mu,\de J^\mu] 
   + g[a_\mu,\de J^\mu]
   + g[a_\mu,\bar J^\mu]
   - g\langle[a_\mu,\de J^\mu]\rangle\ .
\eeq
This is \eq{DdJ}. Here, in addition to \eq{someintegrals}, we made use of
\bea
 2\, g\int dPdQ\, Q\ \Tr\, \big([Q,\partial ^Q]\, a_\mu\big)\,
   \bar f(x,p,Q) &=& g[a_\mu,\bar J^\mu] \ .
\eea
The background covariant derivative of \eq{M-NAJ-2} is
given as
\beq \label{consistency2b}
0=  [\bar D_\mu,\de J^\mu]
  + g\left[a_\nu,[\bar D_\mu,\bar F^{\mu\nu}]\right]
  - [\bar D_\mu,J_{{\mbox{\tiny fluc}}}^\mu ]
  + [\bar D_\mu,\langle J_{{\mbox{\tiny fluc}}}^\mu\rangle ] \ .
\eeq 
Subtracting these equations yields the consistency condition
\beq\label{consistencyF}
0 = [\bar D_\mu,J_{{\mbox{\tiny fluc}}}^\mu ]
   + g[a_\mu,\de J^\mu]
   - [\bar D_\mu,\langle J_{{\mbox{\tiny fluc}}}^\mu\rangle ]
   -g \langle[a_\mu,\de J^\mu]\rangle
   + g[a_\mu,\bar J^\mu]-g\left[a_\nu,[\bar D_\mu,\bar F^{\mu\nu}]\right] \ .
\eeq
Using \eq{M-NAJ-1}, \eq{consistencyM} and \eq{general-cb} we 
confirm \eq{consistencyF}
explicitly. This establishes  the self-consistent conservation of the
fluctuation current.

\subsection{Gauge-consistent approximations}\label{consistentApprox}

We close this section with a comment on the gauge consistency of {\it
approximate} solutions. The consistent current conservation can no longer
be taken for granted when it comes to finding approximate solutions of the
equations. On the other hand, finding an explicit solution will require
some type of approximations to be performed. The relevant question in this
context is to know which approximations will be consistent with gauge
invariance. 

Consistency with gauge invariance requires that approximations have to be
consistent with the background gauge symmetry. From the general discussion
above we can already conclude that dropping any of the explicitly written
terms in \eq{NA-Macro} -- \eq{NA-fluc} is consistent with the background
gauge symmetry \eq{BGS}. This holds in particular for the polarization or
second moment approximation \eq{2nd}.

Consistency of the second moment approximation \eq{2nd} with covariant current
conservation turns out to be more restrictive. Employing $J_{{\mbox{\tiny
fluc}}}=\langle J_{{\mbox{\tiny fluc}}} \rangle$ implies that
\eq{consistencyM} is only satisfied, if in addition
\beq\label{con1}
0=\left[\bar D_\nu,\llangle \left[a_\mu,[a^\mu,a^\nu]\right]\rrangle\right]
\eeq
holds true. This is in accordance with neglecting cubic correlators
for the collision integrals.

Similarly, the consistent conservation of the fluctuation
current implies the consistency condition \eq{consistencyF}, and holds if
\beq\label{con2}
	0=\left[a_\mu,\langle J_{{\mbox{\tiny fluc}}}^\mu \rangle\right]
\eeq
vanishes. It is interesting to note that the consistent current conservation
relates the second moment approximation with the neglection of correlators
of gauge field fluctuations.  We conclude, that \eq{2nd} with \eq{con1} 
and \eq{con2} form a
gauge-consistent set of approximations.

\section{Physical scales: classical versus  quantum plasmas}\label{scales}

In the remaining part of the paper, we study the classical and quantum
non-Abelian plasma close to equilibrium. Prior to this, we shall present a  
discussion of the relevant physical scales
of both relativistic classical and quantum plasmas, close to equilibrium.
We will restore here the fundamental
constants $\hbar, c$ and $k_B$ in the formulas.

To discuss the relevant physical scales in the 
classical non-Abelian plasma, it is convenient to discuss
first the simpler Abelian case, which has been considered in detail in
the literature \cite{K}. At equilibrium  the classical distribution
function is given by the relativistic Maxwell distribution,
\beq
\bar f^{\rm eq} (p_0) = A e^{ - p_0/k_B T} \ ,
\label{eq-class}
\eeq
where $A$ is a dimensionful constant which is fixed once the mean density of
particles in the system is known.  For massless particles $p_0 = p c$,
and if no further internal degrees of freedom are present,  the mean
density is  given by $\bar N = 8 \pi A (k_B T/c)^3$.
The interparticle distance is then $\bar r \sim {\bar N}^{-1/3}$. As we 
are considering a classical plasma, we are assuming
$\bar r \gg \lambda_{\rm dB}$, where $\lambda_{\rm dB}$ is the de
Broglie wave length, $\lambda_{\rm dB} \sim \hbar/p$, 
with $p$ some typical momenta associated to the particles, 
thus $p\sim k_B T/c$.
The previous inequality implies therefore $A \ll 1/\hbar^3$, which is the
condition under which quantum statistical effects can be neglected.

There is another typical scale in a plasma close to equilibrium, 
which is the Debye length
$r_D$. The Debye length is the distance over
which the screening effects of the electric fields in  the
plasma are felt.  For an electromagnetic plasma, the Debye length squared 
is given by \cite{K}
\beq
r^2_D = {k_B T}/{ 4 \pi \bar N  e^2} \ .
\label{ADeb-class}
\eeq
Notice that the electric charge contained in the above formula
is a dimensionful parameter: it is just the electric charge of
the point particles of the system.

In the classical case, and in the absence of the fundamental constant 
$\hbar$, the only  dimensionless quantity that can be constructed 
from the basic scales of the problem is the {\it plasma parameter} 
$\epsilon$. The plasma parameter is defined as the ratio  \cite{K}
\beq
\epsilon = {\bar r^3}/{r^3_D} \ .      \label{pla-par}
\eeq
The quantity $1/\epsilon$ gives the number of particles 
contained in a sphere of radius
$r_D$. If $\epsilon \ll 1$ this implies that 
a large number of particles are in
that sphere, and thus a large number of particles are interacting 
in this volume, and the collective character of their interactions 
in the plasma can not be neglected. 
For the kinetic description to make sense, $\epsilon$ has to be
small \cite{K}. This does not require, in general,  
that the interactions have to be weak and treated perturbatively.

Let us now consider the non-Abelian plasma. The interparticle distance 
is defined as in the previous case. The main differences with respect 
to the Abelian case concerns the Debye length, defined as the distance 
over which the screening effects of the non-Abelian electric fields in  
the plasma are noticed. It reads
\beq
r^2_D = { k_B T}/{ 4 \pi \bar N  g^2 C_2} \ ,
\label{NADeb-class}
\eeq
where   $C_2$, defined in (\ref{quadraticQ}),
is  a dimensionful quantity, carrying the same dimensions as the electric
charge squared in (\ref{ADeb-class}). The coupling constant $g$ 
is a dimensionless parameter. In the non-Abelian plasma one can also 
construct the plasma parameter, defined as in (\ref{pla-par}).

It is interesting to note that there are  two  natural dimensionless 
parameters in the non-Abelian plasma: $\epsilon$ and $g$. 
The condition for the plasma parameter being small translates into
\beq
\left(\frac{4 \pi C_2}{k_B T} \right)^{3/2} \bar N^{1/2} g^3 \ll 1 \ ,
\eeq
which is certainly satisfied for small gauge coupling constant $g \ll 1$. 
But it can also be fulfilled for a rarefied plasma. Thus, one may have 
a small plasma parameter {\it without} having a small gauge coupling 
constant. This is an interesting observation, since the  inequalities 
$\epsilon \ll 1$ and $g \ll 1$ have different physical meanings. A small 
gauge coupling constant allows for treating the non-Abelian interaction 
perturbatively, while $\epsilon \ll 1$ just means having a collective 
field description of the physics occurring in the plasma. In principle, 
these two situations are different. If we knew how to treat the 
non-Abelian interactions {\it exactly}, we could also have a kinetic 
description of the classical non-Abelian plasmas without requiring  
$g \ll 1$.

Now we consider the quantum non-Abelian plasma, and consider the 
quantum counterparts of all the above quantities, as derived from  
quantum field theory. It is common to change the phase space measure 
$d^3x d^3 p$ to the standard quantum normalization 
$d^3 x d^3 p/(2 \pi \hbar)^3$, which makes the quantum distribution
function dimensionless.\footnote{As done in the standard textbooks 
the factor $(2 \pi\hbar)^3$ is introduced into the measure, although 
some authors introduce it into the distribution function. We follow the 
first option.}
For a quantum plasma at equilibrium the one particle distribution 
function is
\beq
\bar f^{\rm eq}_{\rm B} (p_0) =  \frac{1}{e^{p_0/k_B T} - 1} \ ,\quad\quad 
\bar f^{\rm eq}_{\rm F} (p_0) =  \frac{1}{e^{p_0/k_B T} + 1} \ ,
\label{eq-quan0}
\eeq
where the subscript B/F  refers to the bosonic/fermionic statistics.
For a plasma of massless particles the mean density is  
$\bar N \sim (k_B T/\hbar c)^3$. The interparticle distance
$\bar r \sim \bar N^{-1/3}$ becomes of the same order as the de 
Broglie wavelength, which is why quantum statistical effects can not be
neglected in this case.

The value of the Debye mass is obtained from  quantum field theory. 
It depends on the specific quantum statistics of the particles 
and their representation of $SU(N)$. From the quantum Debye mass one 
can deduce the value of the Debye length, which is of order
\beq
r^2 _D \sim \frac{1}{g^2} \left(\frac{\hbar c}{k_B T} \right)^2 \ .
\eeq
It is not difficult  to check that the plasma parameter, 
defined as in (\ref{pla-par}), becomes proportional to $g^3$. 
Thus $\epsilon$ is small {\it if and only if} $g \ll 1$. 
This is so, because in a quantum field theoretical
formulation one does not have the freedom to fix the mean 
density $\bar N$ in an arbitrary way, as in the classical case. This 
explains why the kinetic description of a quantum 
non-Abelian plasma is deeply linked to the small gauge 
coupling regime of the theory.

\section{The Classical Plasma Close to Equilibrium}\label{close2eq}

In this section we put the method to work for a hot non-Abelian plasma 
close to equilibrium. A prerequisite for a kinetic description to be 
viable is a small  plasma parameter $\epsilon\ll 1$. We shall ensure 
this by imposing a small gauge coupling constant $g\ll 1$. Then, all 
further approximations as detailed in the sequel can be seen as a 
systematic expansion in powers of the gauge coupling.

\subsection{Non-Abelian Vlasov equations}

We begin with the set of mean field equations (\ref{NA-Macro}) and 
neglect the effect of statistical fluctuations entirely, 
$\de f\equiv 0$. In that case, (\ref{NA-Macro}) becomes the non-Abelian 
Vlasov equations \cite{Heinz}
\begin{mathletters}\label{NAV-Macro}
\begin{eqnarray}
p^\mu\left(\bar D_\mu - g\, Q_a \bar F^a_{\mu\nu}\, 
\partial _p^\nu\right)\bar f&=& 0 \ ,          \label{NAV-1}
\\
\bar D_\mu  \bar F^{\mu\nu} &=& \bar J^\nu \ , \label{NAVJ-1} 
\end{eqnarray}
where the colour current is given by
\beq
\bar J^\mu_a (x) = 
g \sum_{\hbox{\tiny helicities}\atop\hbox{\tiny species}} 
                   \int dP dQ\, Q_a p^\mu \bar f (x,p,Q) \ ,
\eeq
\end{mathletters}%
We will omit the species and helicity indices on the distribution 
functions, and in the sequel, we will also omit the above sum, in order 
to keep the notation as simple as possible. Equation \eq{NAV-1} is then 
solved perturbatively, as it admits a consistent expansion in powers 
of $g$. Close to equilibrium, we expand the distribution function 
as in \eq{f-g} up to leading order in the coupling constant
\beq
\label{m-close-eq-f}
\bar f(x,p,Q)=     \bar f^{\rm eq}(p_0)
               + g \bar f^{(1)}(x,p,Q) \ .
\eeq
In the strictly classical approach, the relativistic Maxwell 
distribution (\ref{eq-class}) at equilibrium is used. 
Here, we consider only massless particles, or massive particles with
$m \ll T$, such that the masses can be neglected in a first approximation.
We will  consider also internal degrees of freedom,  two helicities
associated to every particle.
 
It is convenient to re-write the equations in terms of current densities.
Consider the current densities
\begin{mathletters}
\begin{eqnarray}
J_{a_1\cdots a_n}^\rho(x,p)
&=& g\ p^\rho  \int dQ \,Q_{a_1}\cdots Q_{a_n} f(x,p,Q), \label{NA-Jp}
\\
{\cal J}_{a_1\cdots a_n}^\rho(x, v)
&=&       \int d\tilde P\, J_{a_1\cdots a_n}^\rho(x,p) \ .\label{NA-Jv}
\end{eqnarray}
\end{mathletters}%
Here, $v^\mu=(1,{\bf v})$ with ${\bf v}^2=1$. The measure $d\tilde P$ 
integrates over the radial components. It is related to \eq{Pmeasure} 
by $dP= d\tilde P d\Omega/4\pi$, and reads
\beq
  d\tilde P = 4\pi\, dp_0\,  d|\vp|\, |\vp|^2\, 2\Theta(p_0)\, \de(p^2) 
\eeq
for massless particles. The colour current  
is obtained performing the remaining angle integration
$J(x)=\int \mbox{\small{$\frac{d\Omega}{4\pi}$}}{\cal J}(x,v)$. 
From now on we will omit the arguments of the
current density $\cal J$, unless necessary to avoid confusion.

We now insert \eq{m-close-eq-f} into \eq{NAV-Macro} and expand in powers
of $g$. The leading order term $p\cdot D \bar f^{\rm eq.}(p_0)$ vanishes.
After multiplying (\ref{NA-1}) by $g Q_a p^\rho/p_0$, 
summing over two helicities, and integrating over $d\tilde P dQ$, we 
obtain for the mean current density at  order $g$
\begin{mathletters}\label{NAV-mean}
\begin{eqnarray} \label{NAV-soft-mean}
v^\mu \bar D_\mu \bar {\cal J}^\rho+ m^2_D v^\rho v^\mu \bar F_{\mu0}
                                &=& 0         \label{NAV-curr}  
\ ,\\
\bar D_\mu\bar F^{\mu\nu}  &=&  \bar J^\nu \ ,\label{soft-YM0}
\end{eqnarray}
\end{mathletters}%
with the Debye mass
\beq
m_D^2 = - 8\pi g^2  C_2 
        \int^\infty_0 dp p^2 \frac{d \bar f^{\rm eq}(p)}{d p} \ . 
\eeq
The solution to  (\ref{NAV-curr}) is now constructed with the
knowledge of the retarded Green's function 
\beq
i v^\mu \bar D_\mu \, G_{{\rm ret}}(x,y;v) = \delta^{(4)} (x-y) \ .
\eeq
It reads
\beq
\label{ret-GF}
G_{{\rm ret}}(x,y;v)_{ab} 
= -i \theta(x_0-y_0)  \delta^{(3)} \left({\bf x}-{\bf y} 
      - {\bf v}(x_0-y_0) \right) \, \bar U_{ab} (x, y) \ ,
\eeq
where  $\bar U_{ab}$ is the parallel transporter obeying   
$v^\mu\bar D_\mu^x \, \bar U_{ab}(x,y)|_{y=x-v t}=0$, and 
$\bar U_{ab}(x,x)= \delta_{ab}$. One finds
\beq
\label{class-HTL-curr}
\bar J^\mu_a (x) =  -m^2_D \int\!
\frac{d\Omega_{{\bf v}}}{4\pi}\,\int^{\infty}_0 d \tau\ 
\bar U_{ab}(x,x-v \tau) v^\mu v^\nu   
\bar F_{\nu 0,b} (x-v\tau)  \ .
\eeq

The above colour current agrees with the hard thermal loop (HTL)
colour current \cite{HTL,KLLM}, except for the value of the Debye mass. 

From \eq{NAV-soft-mean} it is easy to estimate the typical momentum 
scale of the mean fields. If the effects of statistical fluctuations 
are neglected (and as we will see, this is equivalent to neglecting 
collisions), the typical momentum scales associated to the mean current 
and the mean field strength are of the order of the Debye mass $m_D$.  
We will refer to those scales as soft scales. The momentum scales with 
momenta $\ll m_D$ will be referred to as ultra-soft from now on.

\subsection{Leading order dynamics for mean fields and fluctuations}

We now allow for  small statistical fluctuations $\delta f(x,p,Q)$ 
around \eq{m-close-eq-f}, writing
\beq\label{f+df}
f(x,p,Q)=\bar f^{\rm eq.}(p_0)+ g \bar f^{(1)}(x,p,Q) +\delta f(x,p,Q)
\eeq 
and re-write the approximations to (\ref{NA-Macro}) and (\ref{NA-fluc}) 
in terms of current densities and their fluctuations. Note that the 
fluctuations $\delta f(x,p,Q)$ in the close to equilibrium case are already
of the order of $g$. This observation is important for the consistent 
approximation in powers of the gauge coupling. As a consequence, the term
$g \bar f^{(1)}$ in \eq{f+df} will now account for the ultra-soft modes 
for momenta $\ll m_D$. Integrating-out the fluctuations results in an 
effective theory for the latter. 

As before, we obtain the dynamical equation for the mean 
current density at leading order in $g$, after multiplying 
(\ref{NA-1}) by $g Q_a p^\rho/p_0$, summing over two helicities, and 
integrating over $d\tilde P dQ$. The result is
\begin{mathletters}                      \label{NA-fluc-mean}
\begin{eqnarray}                         \label{soft-mean}
  v^\mu \bar D_\mu \bar {\cal J}^\rho
+ m^2_D v^\rho v^\mu \bar F_{\mu0}
& = & \left\langle \eta^\rho\right\rangle 
     +\left\langle \xi^\rho\right\rangle
\ ,\\
    \bar D_\mu\bar F^{\mu\nu} 
   +\left\langle J_{\mbox{\tiny fluc}}^{\nu}\right\rangle 
&=&  \bar J^\nu \ .                       \label{soft-YM}
\end{eqnarray}
\end{mathletters}%
In a systematic expansion in $g$, we have to neglect cubic correlator 
terms as compared to quadratic ones, as they are suppressed explicitly 
by an additional power in $g$. Therefore, we find to leading order 
\begin{mathletters}\label{NA-log-1}
\begin{eqnarray}
\eta^\rho_a  & = &
g \int d \tilde P\, \frac{p^\rho}{p_0} 
 (\bar D_\mu a_\nu-\bar D_\nu a_\mu)^b\,
  \partial^\nu_p \,\delta J^{\mu}_{ab}(x,p)  \ ,        \label{NA-eta1}
\\ 
\xi^\rho_a   & = &
  -gf_{abc}v^\mu\, a^b_{\mu}\ \delta {\cal J}^{c,\rho}\ ,\label{NA-xi1}
\\
 J_{\mbox{\tiny fluc}}^{\rho,a}&=&  
 g f^{dbc}\! \left(\bar D_{\mu}^{ad}  a_{b}^{\mu} a_c^\rho  
+\delta^{ad}  a^b_\mu\! \left(\bar D^\mu a^\rho
-\bar D^\rho a^\mu\right)^c  \right) .                \label{NA-Jfluc1}
\end{eqnarray}
\end{mathletters}%
The same philosophy is applied to the dynamical equations for the 
fluctuations. To leading order in $g$, the result reads
\begin{mathletters}\label{NA-log-2}
\begin{eqnarray}
&&\left(v^\mu \bar D_\mu \,\delta{\cal J}^\rho \right)_a \, 
= - m^2_D v^\rho v^\mu \left(\bar D_\mu a_0-\bar D_0 a_\mu\right)_a 
  - g f_{abc}v^\mu a_\mu ^b \bar {\cal J}^{c,\rho}       \label{vD-dJa} 
\ ,\\
&&v^\mu \left(\partial_\mu\de_{ac}\de_{bd} + g\bar A_\mu^m
\left( f_{amc}\,\delta_{bd} + f_{bmd}\delta_{ac}\right) \right) 
\delta {\cal J}^{\rho}_{cd} 
=  g v^\mu a_\mu^m 
\left( f_{mac}\,\delta_{bd} +f_{mbd}\delta_{ac}\right)   
\bar{\cal J}^{\rho}_{cd}                                \label{vD-dJab}
\ , \\
&&\left(\bar D^2 a^\mu-\bar D^\mu(\bar D a)\right)_a+2 g f_{abc} 
\bar F_b^{\mu\nu}a_{c,\nu}
= \delta J_a^\mu  \ .                                       \label{dJa}
\end{eqnarray}
\end{mathletters}%
The typical momentum scale associated to the fluctuations 
can be estimated from \eq{NA-log-2}. We find that it is of the 
order of the Debye mass $\sim m_D$, that is, of the same order 
as the mean fields in \eq{NAV-mean}. This confirms explicitly the
discussion made above. The typical momentum 
scales associated to the mean fields in \eq{NA-fluc-mean} is therefore 
$\ll m_D$. 

\subsection{Integrating-out the fluctuations}

We solve the equations for the fluctuations (\ref{NA-log-2}) with an 
initial boundary condition for $\delta f$, and $a_\mu(t=0)=0$. Exact 
solutions to (\ref{vD-dJa}) and (\ref{vD-dJab}) can be obtained.

Let us start by solving the homogeneous differential equation
\beq\label{homogeneous}
v^\mu \bar D_\mu \,  \delta   {\cal J}^\rho  = 0 \ ,
\eeq 
with the initial condition $\de {\cal J}^\mu_a(t=0, {\bf x},v)$. 
It is not difficult to check, by direct inspection, that the solution 
to the homogeneous problem is
\beq
\delta {\cal J}^\rho _a (x,v) = \bar U_{ab} (x, x-v t) 
\, \delta {\cal J}^\rho _b(t=0, {\bf x}-{\bf v} t, v) \ .
\eeq
The solution of  (\ref{vD-dJa}) is now constructed using the retarded 
Green's function (\ref{ret-GF}). For $x_0 \equiv t \geq 0$  the 
complete solution can be expressed as 
\begin{eqnarray}
\delta {\cal J}^\rho _a (x,v) &=&  
- \int_0 ^{\infty} d \tau \, \bar U_{ab} (x, x_\tau)
  \left(  \ m^2_D v^\rho v^\mu \left(\bar D_\mu a_0
        - \bar D_0 a_\mu\right)^b (x_\tau) 
        + g f_{bdc}  v^\mu a_\mu ^d (x_\tau)
            \bar {\cal J}^\rho_c(x_\tau, v)  
  \right) 
\nonumber\\[1ex] &&
+\bar U_{ab} (x, x_t) \, 
\delta {\cal J}^\rho _b (x_t, v)\,.               \label{dJ-expl} 
\end{eqnarray}
We have  introduced $x_\tau \equiv x - v \tau$, and thus 
$x_t=(0, {\bf x}-{\bf v} t)$. Since $a_\mu(t=0)=0$, one can check that
the above current obeys the correct initial condition. 

The equation (\ref{vD-dJab}) can be solved in a  similar way. 
The solution is 
\begin{eqnarray}
\delta {\cal J}^\rho _{ab} (x,v) & = & 
\bar U_{am} (x, x_t) \bar U_{bn} (x, x_t)
\, \delta {\cal J}^\rho _{mn} (x_t, v) 
\nonumber\\ 
&& - g\int_0 ^{\infty}d\tau \,
\bar U_{am} (x, x_\tau)
\bar U_{bn} (x, x_\tau)
\left(f_{mpc} \delta_{nd}  +  f_{npd} \delta_{mc}\right)
v^\mu a_\mu ^p (x_\tau)\bar {\cal J}^\rho_{cd}(x_\tau, v) \ . 
\end{eqnarray}
Now we seek for solutions to the equation (\ref{dJa}) with the colour
current of the fluctuation as  found above.
However, notice that this equation is non-local in $a_\mu$, 
which makes it difficult
to find exact solutions. Nevertheless, one can solve the equation in
an iterative way, by making  a double expansion 
in both $g\bar A$ and $g\bar {\cal J}$. This is possible since 
the parallel transporter 
$\bar U$ admits an expansion in $g \bar A$, so that the current 
$\delta {\cal J}^\rho$ can be expressed as a power series in $g \bar A$
\beq
\delta {\cal J}^\rho =  \delta {\cal J}^{\rho (0)} 
                       +\delta {\cal J}^{\rho (1)} 
                       +\delta {\cal J}^{\rho (2)} + \cdots \ ,
\eeq
and thus  (\ref{dJa}) can be solved for every order in $g \bar A$. To 
lowest order in $g \bar A$, 
using $\bar U_{ab} = \delta_{ab} + {\cal O}(g \bar A)$,  
equation (\ref{dJa}) becomes
\beq
  \partial^\mu \left( \partial_\mu a_{\nu, a}^{(0)} 
- \partial_\nu a_{\mu, a}^{(0)} \right) 
= \delta J_{\nu, a}^{(0)} \ .                       \label{Abelgeq} 
\eeq
Using the one-sided Fourier transform \cite{K}, and \eq{dJ-expl}, we find
\begin{eqnarray}
\delta  J _{a\,+}^{\mu \, (0)}(k)
&=& \Pi^{\mu \nu}_{ab}(k) a_{\nu,b}^{(0)} (k)
    - g f_{abc} \int\!\!\frac{d\Omega_{\bf v}}{4\pi}
        \frac{1}{-i\ k \cdot v} \int\frac{d^4q}{(2\pi)^4} 
        v^\rho a_{\rho}^{b(0)}(q) \, \bar{\cal J}^{\mu,c}(k-q,v) 
\nonumber \\ &&  
+\int\frac{d\Omega_{\bf v}}{4\pi}
 \frac{\delta{\cal J}^{\mu}_{a}(t=0,{\bf k},v)}{-i\, k\cdot v} \ ,
                                                             \label{dJ+}
\end{eqnarray}
where $\Pi^{\mu \nu}_{ab}(k)$ is the polarization tensor in the plasma,
which reads
\beq
\Pi^{\mu \nu}_{ab}(k) 
= \delta_{ab} m^2_D 
    \left( - g^{\mu 0} g^{\nu 0} 
           + k_0 \int\!\!\frac{d\Omega_{\bf v}}{4\pi}
                         \frac{v^\mu v^\nu}{k \cdot v} 
    \right) \ ,
\eeq
and agrees with the HTL polarization tensor of QCD \cite{HTL,KLLM}, 
except in the value of the Debye mass.
Retarded boundary conditions are assumed above, with the 
prescription $k_0 \rightarrow k_0 + i 0^+$.

We solve (\ref{Abelgeq}) iteratively  in momentum space for $a_\mu$
as an  infinite power series in $g\bar {\cal J}$,
\beq
a_\mu^{(0)}= a_\mu^{(0,0)}
            +a_\mu^{(0,1)}
            +a_\mu^{(0,2)}+\ldots
\eeq
where the second index counts the powers of the background 
current $g\bar {\cal J}$.
Notice that in this type of Abelianized approximation, the equation 
(\ref{Abelgeq}) has a (perturbative) Abelian gauge symmetry associated 
to the fluctuation $a_\mu$. This symmetry is only broken by the term 
proportional to $\bar {\cal J}$ in the current. It is an exact 
symmetry for the term $a_\mu ^{(0,0)}$ in the above expansion.  We 
will use this perturbative gauge symmetry in order to simplify the 
computations, and finally check that the results of the approximate 
collision integrals do not depend on the choice of the fluctuation gauge.
  
Using the one-sided Fourier transform, we find
the following results for the longitudinal fields, 
in the gauge ${\bf k} \cdot {\bf a}^{(0,0)} =0$, 
\begin{mathletters}                                         \label{aL+}
\begin{eqnarray}
a_{0,a\,+}^{(0,0)}(k)
&=& \frac{1}{{\bf k}^2-\Pi_L}
    \int\!\!\frac{d\Omega_{\bf v}}{4\pi}
     \frac{\delta{\cal J}_{0,a}(t=0,{\bf k},v)}{-i\ k \cdot v}\ , 
                                                            \label{aL0}
\\
a_{0,a\,+}^{(0,1)}(k)
&=& \frac{-g f_{abc}}{{\bf k}^2-\Pi_L}
    \int\frac{d\Omega_{\bf v}}{4\pi}\frac{1}{-i\ k\cdot v}
    \int\frac{d^4q}{(2\pi)^4} v^\mu a_{\mu}^{b(0,0)}(q) \, 
    \bar{\cal J}_0^c(k-q,v) \ ,                             \label{aL1}
\end{eqnarray}
\end{mathletters}%
while we find
\begin{mathletters}                                         \label{aT+}
\begin{eqnarray}
a_{i,a\,+}^{T(0,0)}(k)
&=&\frac{1}{-k^2+\Pi_T}
   \int\!\!\frac{d\Omega_{\bf v}}{4\pi}
   \frac{\delta{\cal J}^{T}_{i,a}(t=0,{\bf k},v)}{-i\ k \cdot v}\ , 
                                                            \label{aT0}
\\
a_{i,a\,+}^{T(0,1)}(k)
&=& \frac{-g f_{abc}}{-k^2+\Pi_T}P_{ij}^T({\bf k})
    \int\frac{d\Omega_{\bf v}}{4\pi}\frac{1}{-i\ k\cdot v}
    \int\frac{d^4q}{(2\pi)^4} v^\mu a_{\mu}^{b(0,0)}(q) \, 
     \bar{\cal J}_j^c(k-q,v) \ ,                            \label{aT1}
\end{eqnarray}
\end{mathletters}%
for the transverse fields.\footnote{In \cite{LM}, we used a more 
condensed notation. There, the functions $a_{i,\,+}^{(0,0)}$ and 
$a_{i,\,+}^{(0,1)}$ have been denoted $a_{i,\,+}^{(0)}$ and
$a_{i,\,+}^{(1)}$.} 
The functions $\Pi_{L/T}(k)$ are the longitudinal/transverse 
polarization tensor of the plasma, 
$P_{ij}^T({\bf k})=\delta_{ij}-{k_ik_j}/{{\bf k}^2}$ the 
transverse projector, and $a_i^T\equiv P^T_{ij}a_j$. 

In the approximation $g \ll 1$, it will be enough to consider
the solution of leading (zeroth) order in $g \bar A$, and the 
zeroth and first order in $g\bar {\cal J}$. The remaining terms 
are subleading in the leading logarithmic approximation.
However, notice that we have all the tools necessary to compute the 
complete (perturbative) series. If we could solve equation (\ref{dJa}) 
exactly, it would not be necessary to use this perturbative expansion.

\subsection{The statistical correlator of fluctuations}

With the explicit expressions obtained in \eq{dJ+}, \eq{aL+} and \eq{aT+}, 
we can express all fluctuations in terms of initial conditions 
$\de {\cal J}^\mu_a(t=0, {\bf x},v)$ and the mean fields. 

In order to compute the correlator of initial conditions, we will make 
use of the result obtained in sect.~\ref{stati-sec}. For each species 
of particles or internal degree of freedom, the statistical average 
over initial conditions can be expressed as
\bea
\langle \delta f(t=0,{\bf x},p,Q)    \, 
        \delta f(t=0,{\bf x}',p',Q') \rangle 
&=& \delta^{(3)}({\bf x}-{\bf x}')
    \delta^{(3)}({\bf p}-{\bf p}') 
    \delta (Q- Q') \bar f(x,p,Q)
\nonumber\\ &&
  + {\tilde g}_2({\bf x},p,Q;{\bf x}',p',Q')\ ,     \label{average}
\eea
where the function ${\tilde g}_2$ obtains from the two-particle 
correlator, and
\beq
\delta (Q -Q') = \frac{1}{c_R} \de({\bphi} - {\bphi}') \
                               \de ({\bpi} -  {\bpi}) \ ,
\eeq
and $\bphi$, ${\bpi}$ are the Darboux variables associated to the
colour charges $Q_a$.
The appearance of the factor $1/c_R$ in the above expression is due to
the change of normalization factors associated to the functions
${\cal N}$ and $f$, as we mentioned at the end of sect.~\ref{stati-sec}.
The above statistical average is all we need to evaluate the 
collision integrals below.

From \eq{average} one deduces the statistical average
over colour current densities $\de {\cal J}$. We expand the 
momentum $\delta$-function  in polar coordinates
\beq
\delta^{(3)}({\bf p}-{\bf p}') 
= \frac{1}{p^2} \,\delta(p - p') \, 
  \delta^{(2)}(\Omega_{\bf v }-\Omega_{\bf v'}) \ ,
\eeq
where $\Omega_{\bf v}$ represents the angular variables associated to 
the vector ${\bf v}= {\bf p} /|{\bf p}|$. After simple integrations 
we arrive at 
\bea
\llangle\delta {\cal J}_\mu^a (t=0,{\bf x},v) \, 
        \delta {\cal J}_\nu^b (t=0,{\bf x}',v')\rrangle 
&=& 2 g^2 B_C \,C_2 \,\delta^{ab} \, v_\mu v'_\nu \, 
     \delta^{(3)}({\bf x}-{\bf x}') \,
     \delta^{(2)} (\Omega_{\bf v }-\Omega_{\bf v'})
\nonumber\\ && 
+\,{\tilde g}^{ab}_{2,\mu\nu}({\bf x},v;{\bf x}',v')\ , \label{cur-cor}
\eea
where $v^\mu= (1,\vv)$, and 
\beq                                                \label{class-f-cor}
B_C = 16\pi^2\int_0^\infty dp\, p^2\,\bar f^{\rm eq}(p) \ .
\eeq
The function ${\tilde g}^{ab}_{2,\mu\nu}$ is obtained from the 
two-particle correlation function $\tilde g_2$. Notice that we have 
neglected the piece $g \bar f^{(1)}$ above, as this is subleading in 
an expansion in $g$.

Since we know the dynamical evolution of all fluctuations we can also
deduce the dynamical evolution of the correlators of fluctuations,
with the initial condition (\ref{average}). This corresponds to
solving \eq{quad-corr} in the present approximation. 
It is convenient to proceed as follows \cite{K}.
We separate the colour current (\ref{dJ-expl}) into a source part 
and an induced part, 
\beq
\delta {\cal J}^\mu =   \delta {\cal J}^\mu _{\mbox{\tiny source}} 
                      + \delta {\cal J}^\mu _{\mbox{\tiny induced}} \ . 
\eeq
The  induced piece $\delta {\cal J}^\mu_{\mbox{\tiny induced}}$ is the
part of the current which contains the dependence on $a_\mu$, and thus 
takes the polarization effects of the plasma into account.
The  source piece $\delta {\cal J}^\mu _{\mbox{\tiny source}}$ is the 
part of the current which depends only on the initial condition, given 
by the solution of the homogeneous equation \eq{homogeneous}. This 
splitting will be useful since ultimately all the relevant correlators 
can be expressed in terms of correlators of 
$\de {\cal J}^\mu_{{\mbox{\tiny source}}}$.

From the explicit solution (\ref{dJ-expl}) and the average (\ref{cur-cor}) 
we then find, at leading order in $g$ and neglecting the non-local term 
in (\ref{cur-cor})
\begin{eqnarray}
 \llangle\delta {\cal J}_{\mbox{\tiny source}}^{a,\mu} (x,v) \ 
         \delta {\cal J}_{\mbox{\tiny source}}^{b,\nu} (x',v')\rrangle 
&=&
2 g^2 B_C \,C_2  v^\mu v'^\nu 
    \delta^{(3)}({\bf x}-{\bf x}'-{\bf v}(t-t'))
    \delta^{(2)} (\Omega_{\bf v }-\Omega_{\bf v'}) 
\nonumber\\ &&\qquad 
    \times \bar U^{ac}(x, x-vt) \bar U^{bc}(x', x'-v't') \ .
\end{eqnarray}
Here, and from now on,  we neglect the non-local piece  
${\tilde g}^{ab}_{2,\mu\nu}$.
It can be shown  \cite{K} that they give contributions which decrease 
rapidly with time, so that for asymptotic large times, they vanish.

Expanding the parallel transporter $\bar U$, and switching to momentum 
space we find the spectral density to zeroth order in $ g \bar A$
\beq
\llangle \delta {\cal J}_{\mu}^a \ 
         \delta {\cal J}_{\nu}^b
\rrangle^{{\mbox{\tiny source}}\,(0)}_{k,v,v'} 
 = 2 g^2 B_C \,C_2 \delta^{ab} v_\mu v'_\nu 
   \delta^{(2)} (\Omega_{\bf v }-\Omega_{\bf v'})
   (2 \pi)\delta (k \cdot v) \ .                    \label{spectral-dJ}
\eeq

As an illustrative example, let us compute the correlator of two transverse 
fields $a$. Using  (\ref{aT0}) and (\ref{spectral-dJ}) one arrives at
\beq                                                
 \llangle a_{i,a}^{T(0,0)}(k)\,                     \label{basic-corr}
          a_{j,b}^{T(0,0)}(q) \rrangle 
=  g^2 B_C \,C_2 \delta_{ab} 
  (2 \pi )^4 \delta^{(4)} (k+q)
  \frac{P_{ik}^T({\bf k})P_{jl}^T({\bf k})}{|-k^2+\Pi_T|^2}  
  \int\!\!\frac{d\Omega_{\bf v}}{4\pi} v_k v_l \,
  \delta (k \cdot v)  \ .  
\eeq
Since the imaginary part of the polarization tensor, 
which describes Landau damping, can be expressed as \cite{KLLM}
\beq
{\rm Im}\, \Pi^{\mu \nu}_{ab} (k) 
= -\delta_{ab} m^2_D \pi  k_0 
   \int \!\!\frac{d\Omega_{\bf v}}{4\pi} 
    v^\mu v^\nu \,\delta (k \cdot v) \ ,            \label{imag}
\eeq
the statistical correlator  can finally be written as
\beq
 \llangle a_{i,a}^{T(0,0)}(k)\,  
          a_{j,b}^{T(0,0)}(q) \rrangle 
= \frac{4 \pi T}{k_0} 
  \frac{{\rm Im}\, \Pi_{ij,T}^{ab}(k)}{|-k^2+\Pi_T|^2}  
  (2 \pi)^3  \delta^{(4)} (k+q) \ .                 \label{first-FDT}
\eeq
Here, we have used the relation
\beq
{2 g^2 C_2 B_C}= 4 \pi T {m^2_D} \ .              \label{magical-rel}
\eeq
Equation (\ref{first-FDT}) is a form of the fluctuation dissipation 
theorem (FDT), which links the dissipative
processes occurring in the plasma with statistical fluctuations.

\subsection{The collision integral}

We are now ready to compute at leading order in $g$ the collision 
integrals appearing on the r.h.s. of \eq{soft-mean}. 
We shall combine the expansions introduced earlier to expand the 
collision integrals in powers of $\bar {\cal J}$ (while
retaining only the zeroth order in $g\bar A$),  
\beq
\langle \xi\rangle=\langle \xi^{(0)}\rangle
+\langle \xi^{(1)}\rangle+\langle \xi^{(2)}\rangle+\ldots\ ,
\eeq
and similarly for $\langle \eta\rangle$  and
$\langle J_{\mbox{\tiny fluc}}\rangle$. We  find that the induced 
current $\langle J_{\mbox{\tiny fluc}}^{(0)}\rangle$
vanishes, as do the fluctuation integrals $\langle \eta^{(0)}\rangle$ 
and $\langle \xi^{(0)}\rangle$.  The vanishing of 
$\langle J_{\mbox{\tiny fluc}}^{(0)}\rangle$
is deduced trivially from the fact that 
$\langle a_a^{(0,0)}  a_b^{(0,0)} \rangle \sim \delta_{ab}$,
while this correlator always appears contracted with the 
antisymmetric constants $f_{abc}$ in
$J_{\mbox{\tiny fluc}}$.  To check that $\langle \eta^{(0)}\rangle = 0$, 
one needs the statistical correlator
$\langle \delta J^\mu _a \delta J^\rho _{ab}\rangle$, which is 
proportional to $\sum_a d_{aab}=0$ for $SU(N)$. 
The vanishing of $\langle \eta^{(0)}\rangle$
is consistent with the fact 
that in the Abelian limit the counterpart of  $\langle \eta\rangle$
vanishes at equilibrium \cite{K}. 
Finally, $\langle \xi^{(0)}\rangle = 0$ due to a
contraction of $f_{abc}$ with a correlator symmetric in the colour indices.

In the same spirit we evaluate the terms in the collision integrals 
containing one 
$\bar{\cal J}$ field and no background gauge $\bar A$ fields. Consider
\begin{eqnarray}
\left\langle\xi^{(1)}_{\rho,a}\right\rangle 
&=& g f_{abc} v^\mu \bigg\{ 
-\left\langle a^{(0,1)}_{\mu,b} (x)\,
        \delta {\cal J}_{\rho,c}^{(0)}(x,v) \right\rangle 
\nonumber \\
&&\qquad\qquad
  + \, g f_{cde} v^\nu  \int^{\infty} _{0} \!\!\! d \tau 
   \bar{\cal J}_{\rho,e}( x_\tau, v) 
   \left\langle   a^{(0,0)}_{\mu,b} (x)\ a^{(0,0)}_{\nu,d} (x_\tau) 
\right\rangle \bigg\}. 
\end{eqnarray}
Using the values for $a_\mu$ and $\delta {\cal J}^{(0)}$ as 
found earlier, we obtain in momentum space
\begin{eqnarray}
&&\left\langle\xi^{(1)}_{\rho,a}(k,v)\right\rangle \approx
-g^4 C_2 N B_C v_\rho
\!\int\!\frac{d\Omega_{{\bf v}'}}{4\pi}\,C({\bf v}, {\bf v}') 
\left( \bar {\cal J}^{0}_a (k,v) -  \bar {\cal J}^{0}_a (k,v') \right) \ , 
\end{eqnarray}
with
\beq\label{C}
C({\bf v}, {\bf v}') = \int \frac{d^4 q}{(2 \pi)^4} 
\left|\frac{v_i P_{ij}^T(q) v'_j}{-q^2+\Pi_T}\right|^2 (2 \pi) 
\delta(q \cdot v) (2 \pi) \delta(q \cdot v')  \ .
\eeq
Here, the symbol $\approx$ means that only the leading terms have 
been retained.
To arrive at the above expression we have used the $SU(N)$ relation 
$f_{abc} f_{abd} = N
\delta_{cd}$. Within the momentum integral, we have 
neglected in the momenta of the mean fields, $k$, in front of the momenta
of the fluctuations, $q$. As we discussed above, the momenta associated 
to the background fields are much smaller than those associated to the 
fluctuations which justifies this approximation to leading order. 
This is precisely what makes the collision integral, which in principle
contains a convolution over momenta, local in $k$-space 
(resp. $x$-space). The only remaining non-locality stems from 
the angle convolution of \eq{C}.  
Notice that we have only written the part arising from  the transverse 
fields $a$, as the one associated to the longitudinal modes
is subleading. This is easy to see once one
realizes that the above integral is logarithmic 
divergent in the  infrared (IR)  region, 
while the longitudinal contribution is finite. At this point, we can 
also note that the collision integral computed this way
is independent on the (perturbative) Abelian gauge used to solve  
equation (\ref{Abelgeq}).
This is so because the collision integral computed this way 
can always be expressed
in terms of the imaginary parts of the polarization 
tensors (\ref{imag}) in the plasma,
which are known to be gauge-independent.
 
The integral \eq{C} has also been obtained in \cite{Arnold}, on the basis of
a phenomenological derivation of the Boltzmann collision integral
 for a quantum plasma. The only difference consists in the value of the 
Debye mass appearing in the polarization tensor. 

In any case, the transverse polarization tensor $\Pi_T$ 
vanishes at $q_0 =0$, and the dynamical
screening is not enough to make \eq{C}  finite. 
 An IR  cutoff must be introduced by hand in order to evaluate the integral.
With a cutoff of order $g m_D$ we thus find at logarithmic accuracy
\beq
C({\bf v}, {\bf v}') \approx \frac{2}{\pi^2 m^2_D}  \ln\left(1/g \right) 
\frac{({\bf v}\cdot{\bf v}')^2}{\sqrt{1-({\bf v}\cdot{\bf v}')^2}}
\eeq
Using also the relation (\ref{magical-rel}) we finally arrive at the 
collision integral to leading logarithmic accuracy,
\begin{eqnarray}
\label{leading-log}
\left\langle \xi^{(1)}_{\rho,a}(x,v)\right\rangle &=&
 -  \frac{g^2}{4\pi} N T \ln\left(1/g \right) 
v_\rho\!\int\!
\frac{d\Omega_{{\bf v}'}}{4\pi}\, {\cal I}(v,v')\bar{\cal J}^0_a(x,v'), \\
\label{IK}
{\cal I}(\vv,\vv') &=&  \delta^{(2)}({\bf v}-{\bf v}') 
-{\cal K}(\vv,\vv')\ ,\quad\quad {\cal K}(\vv,\vv')=\frac{4}{\pi}
\frac{({\bf v}\cdot{\bf v}')^2}{\sqrt{1-({\bf v}\cdot{\bf v}')^2}} \ ,
\end{eqnarray}
where we have introduced 
$ \delta^{(2)}({\bf v}-{\bf v}') 
\equiv 4 \pi  \delta^{(2)}(\Omega_{\bf v}- \Omega_{\bf v}') $, 
$\int \frac{d\Omega_\vv}{4 \pi} \delta^{(2)}({\bf v}-{\bf v}')=1$.

We can verify explicitly that the collision integral to leading 
logarithmic accuracy is consistent with gauge invariance. This should 
be so, as the approximations employed have been shown in 
sect.~\ref{consistentApprox} on general grounds to be consistent with 
gauge invariance. Evaluating the correlator in (\ref{DJ}) in the 
leading logarithmic approximation yields 
\beq\label{c-check}
g f_{abc} \left\langle a^b_\mu (x) \de J^\mu_c (x) \right\rangle 
= - \frac{g^2}{4\pi} N T \ln\left(1/g \right) 
    \int \frac{d\Omega_{{\bf v}}}{4\pi}\,
         \frac{d\Omega_{{\bf v}'}}{4\pi}\, 
         {\cal I}(\vv,\vv')\bar{\cal J}^0_a(x,v') \ ,
\eeq
which vanishes, because 
\beq\label{averageI}
\!\int\! \frac{d\Omega_{{\bf v}}}{4\pi}\, {\cal I}(\vv,\vv') = 0 \ .
\eeq
We thus establish that $\bar D_\mu \bar J^\mu =0$, in accordance 
with (\ref{soft-YM}) in the present approximation.

\subsection{The source for stochastic noise}

The collision integral obtained above describes a dissipative process in 
the plasma, so in principle, it could trigger the system to abandon 
equilibrium. Whenever dissipative processes are encountered, it is 
important to identify as well the stochastic source related to it. This 
is the essence of the fluctuation-dissipation theorem (FDT). 
Phenomenologically, this is well known, and sometimes used the other way 
around: imposing the FDT allows to add by hand a source for stochastic 
noise with the strength of its self-correlator fixed by the dissipative 
processes. 

In the present formalism, we are able to identify directly the source
for stochastic noise which prevents the system from abandoning equilibrium. 
This proves, that the FDT does hold (analogous considerations have been 
presented in \cite{Bodeker}).
The relevant noise term is given by the contributions from the transversal
gauge fields in $\xi^{(0)}$. While its average vanishes,
$\langle \xi^{(0)} \rangle=0$, its correlator 
\beq
\llangle \xi^{\rho (0)}_a (x,v)\, \xi^{\sigma (0)}_b (y,v') \rrangle 
= g^2 f_{apc} f_{bde} v^\mu v'^\nu
  \llangle 
  a^p_\mu (x)\,\delta {\cal J}_{\mbox{\tiny source}}^{\rho, c}(x,v)\, 
  a^d_\nu (y)\,\delta {\cal J}_{\mbox{\tiny source}}^{\sigma, e}(y,v')
  \rrangle^{(0)}
\eeq
does not.  In order to evaluate this correlator we switch to
Fourier space. Within the second moment approximation we expand the
correlator $\langle \delta f \delta f \delta f \delta f \rangle$
into products of second order correlators 
$\langle \delta f \delta f\rangle \langle \delta f \delta f \rangle$
and find  
\begin{eqnarray}
\llangle \xi^{\rho (0)}_a (k,v)\,  \xi^{\sigma (0)}_b (p,v') \rrangle 
& = &  g^2 f_{apc} f_{bde} v^\mu v'^\nu \int \frac{d^4 q}{(2 \pi)^4} 
      \int \frac{d^4 r}{(2 \pi)^4} 
\nonumber \\
&& \left \{\ \, 
\llangle a^{(0,0)}_{\mu p} (q)\,a^{(0,0)}_{\nu d} (r) \rrangle\,  
\llangle \delta {\cal J}_{\mbox{\tiny source}}^{(0) \rho, c} (k-q,v)\, 
         \delta {\cal J}_{\mbox{\tiny source}}^{(0) \sigma, e}(p-r,v') 
\rrangle \right. \nonumber 
\\ && + \left. 
\llangle a^{(0,0)}_{\mu p} (q)\,
         \delta {\cal J}_{\mbox{\tiny source}}^{(0)\sigma, e} (p-r,v') 
\rrangle\,  
\llangle \delta {\cal J}_{\mbox{\tiny source}}^{(0) \rho, c} (k-q,v)\, 
         a^{(0,0)}_{\nu d}(r) 
\rrangle 
\right \}
\end{eqnarray}
In the leading logarithmic approximation we retain only the contributions 
from the transverse modes. Evaluating the correlators leads to
\beq
\llangle \xi^{\mu,a}_{(0)} (x,v) \,
         \xi^{\nu,b}_{(0)} (y,v') \rrangle 
= \frac{g^6  N C_2^2  B^2_C} {(2 \pi)^3 m^2_D}\, 
  \ln{(1/g)}\, v^\mu v'^\nu\, {\cal I}(\vv,\vv')\, 
  \delta^{ab}\, \delta^{(4)} (x-y) \ .               \label{noi-Bolt-0}
\eeq
After averaging over the angles of ${\bf v}$ and ${\bf v'}$, and 
using the relation (\ref{magical-rel}), the correlator becomes
\beq
\llangle \xi^{i,a}_{(0)} (x)\, \xi^{j,b}_{(0)}  (y) \rrangle 
= 2 T\, \frac{m^2_D}{3}\, \frac{g^2}{4 \pi}\, N T \ln{(1/g)}\, 
  \delta^{ab}\, \delta^{ij} \,\delta^{(4)} (x-y) \ .   \label{noi-Bolt}
\eeq
In particular, all correlators 
$\langle \xi^{0}_{(0)} (x) \xi^{\mu}_{(0)}(y) \rangle$ vanish. 
Equation \eq{noi-Bolt} identifies $\xi_{(0)}^i(x)$ as a source of 
white noise. 

The presence of this noise term does not interfere with the covariant 
current conservation confirmed at the end of the previous section. 
This can be seen as follows. The noise term enters \eq{c-check} as the 
angle average over the $0$-component of $\xi^\mu_{(0)}(x,v)$. As we 
have established above, the logarithmically enhanced contribution from 
the noise source stems from its correlator \eq{noi-Bolt-0}. Averaging 
the temporal component of \eq{noi-Bolt-0} over the angles of $\bf v$, 
and using \eq{averageI}, it follows that
\beq
\llangle
        \xi_{(0)}^{0,a}(x,v' )
        \int \frac{d\Omega_\vv}{4\pi}\,\xi^{0,b}_{(0)}(y,v)
\rrangle
=0  \ .                                               \label{no-noise}
\eeq 
We thus conclude, that the temporal component of the noise,
$\xi^0_{(0)}(x,v)$, has no preferred $\vv$-direction, which entails that 
$\int \frac{d\Omega}{4\pi}\xi^0_{(0)}(x,v)=0$  
in the leading logarithmic approximation. Thus, the mean current 
conservation is not affected by the noise term.

\subsection{Mean Field Equations and Non-Abelian Ohm's law}
\label{mean}

We have managed to obtain the following set of mean field equations,
after integrating-out the statistical fluctuations (from now on, we
drop the bar to denote the mean fields)
\begin{mathletters} \label{fin-set}
\begin{eqnarray} 
v^\mu  D_\mu {\cal J}^\rho(x,v)
&=& - m^2_D v^\rho v^\mu  F_{\mu0} (x)
    - \gamma\  v^\rho\!\int\! \frac{d\Omega_{{\bf v}'}}{4\pi}\, 
      {\cal I}(\vv,\vv') {\cal J}^0(x,v') 
    + \zeta^\rho(x,v)\, ,                               \label{last-Bol}
\\ 
D_\mu F^{\mu\nu}  
&=& \int\!\frac{d\Omega_{{\bf v}}}{4\pi}\,{\cal J}^\nu(x,v)\, .
\end{eqnarray}
\end{mathletters}%
Here, we denote by $\zeta^\rho(x,v)$ the stochastic noise term identified 
in the preceding section, its correlator given by \eq{noi-Bolt-0}. 
We also introduced 
\beq  \gamma = \frac{g^2}{4\pi} N T \ln\left(1/g \right)\ ,\eeq 
which will be identified as (twice) the damping rate for the 
ultra-soft currents. We 
shall refer to \eq{last-Bol} as a Boltzmann-Langevin equation, as it 
accounts for particle interactions via a collision integral as well 
as for the stochastic character of the underlying fluctuations.

The Boltzmann-Langevin equation \eq{last-Bol} has three distinct scale 
parameters, the temperature $T$, the Debye mass $m_D$, and the damping term 
$\gamma$. In the leading logarithmic approximation, these scales are 
well separated, $\gamma\ll m_D\ll T$. This is why \eq{fin-set} is 
dominated by different terms, depending on the momentum range considered. 
For hard momenta, \eq{last-Bol} is only dominated by the l.h.s., reducing 
it to the (trivial) current of hard particles moving on world lines. 
For momenta about the Debye mass, the term proportional to $m_D^2$ 
becomes equally important, while the noise term and the collision integral 
remain suppressed by $\gamma/m_D$. The resulting current is then given 
by \eq{class-HTL-curr}, the HTL current. Momentum modes below the Debye 
mass are affected by the damping term of the collision integral. 
Close to the scale of the Debye mass, the higher order corrections 
to \eq{class-HTL-curr} are obtained as an expansion in 
$\gamma/v\cdot D$, 
\beq
{\cal J}^\mu(x,v) 
= \sum_{n=0}^{\infty}  {\cal J}^\mu_{(n)}(x,v) \ , \label{seriesJ}
\eeq
where the current densities ${\cal J}^\mu_{(n)}(x,v)$ obey the 
differential equations 
\begin{mathletters}\label{seriesA}
\begin{eqnarray} 
v^\mu  D_\mu {\cal J}^\nu_{(0)}(x,v)
  &=& - m^2_D  v^\nu v^\mu  F_{\mu0} (x)  + \zeta^\nu(x,v) 
\\
v^\mu  D_\mu {\cal J}^\nu_{(n)}(x,v)
  &=&- \gamma  v^\nu \int\! \frac{d\Omega_{{\bf v}'}}{4\pi}\, 
                 {\cal I}(\vv,\vv') {\cal J}^0_{(n-1)}(x,v') 
\eea
\end{mathletters}%
Apart from the noise term, the leading order term in this expansion, 
${\cal J}^\nu_{(0)}(x,v)$, coincides with the HTL current 
\eq{class-HTL-curr}. All higher order terms 
${\cal J}^\mu_{(n)}(x,v)$ are smaller by powers of 
$\sim (\gamma/v\cdot D)^n$, and recursively given by
\begin{mathletters}\label{solutionA}
\bea
{\cal J}^\nu_{(0)}(x,v) 
&=& \int^{\infty}_0 d \tau U(x,x-v \tau) 
    \left\{ -m^2_D v^\nu v^\mu  F_{\mu0} (x-v\tau) 
            + \zeta^\nu(x-v\tau,v) \right\} \ ,
\\
{\cal J}^{\nu}_{(n)} (x,v) 
&=& -\gamma\,\int^{\infty}_0 d \tau
       U(x,x-v \tau)  v^\nu \int\! \frac{d\Omega_{{\bf v}'}}{4\pi}\, 
                 {\cal I}(\vv,\vv') {\cal J}^0_{(n-1)}(x,v') 
\eea
\end{mathletters}%
This expansion is consistent with covariant current conservation. 
For every partial sum up to order $n$, we have
\beq
  D_\mu \left(  J^\mu_{(0)}
              + J^\mu_{(1)}
              + \ldots 
              + J^\mu_{(n)} \right) = 0
\eeq
This expansion describes correctly how the presence of the collision 
integral modifies the ultra-soft current. However, it has 
a limited domain of validity. In particular, the overdamped
regime where $v\cdot D\ll \gamma$ can not be reached, because the effective
expansion parameter diverges.  

Alternatively, one can separate the local from the non-local part
of the collision integral to perform an expansion in the latter only. 
The effective expansion parameter is then $\gamma/(v\cdot D + \gamma)$,
which has a better IR behaviour. We expect this expansion to be
much better for the spatial than for the temporal component of 
${\cal J}^\rho(x,v)$. This is so, because the term proportional to 
the non-local part ${\cal K}(\vv,\vv')$ of the collision integral 
in \eq{last-Bol} gives no contribution to the dynamical equations of 
the spatial component $J^i$, after angle averaging \eq{last-Bol}
over the directions of $\vv$. However, for the dynamical equation of the
temporal component, this term precisely cancels the local damping term,
which is of course a direct consequence of current conservation.

In this light, we decompose the current as in \eq{seriesJ} to find 
the differential equations 
\begin{mathletters}\label{seriesB}
\bea
(v^\mu  D_\mu +\gamma) {\cal J}^\rho_{(0)}(x,v)
&=& - m^2_D v^\rho v^\mu  F_{\mu0} (x) + \zeta^\rho(x,v)\ , 
\label{soft-current-B}
\\
(v^\mu  D_\mu+\gamma) {\cal J}^\rho _{(n)}(x,v)
&=& \gamma\  v^\rho\int\!\frac{d\Omega_{{\bf v}'}}{4\pi}\, 
          {\cal K}(\vv,\vv') {\cal J}^0_{(n-1)}(x,v') .
\end{eqnarray}
\end{mathletters}%
The retarded Green's function of the differential operator
\beq
i \left(v^\mu D_\mu + \gamma  \right) \, G_{{\rm ret}}(x,y;v) 
= \delta^{(4)} (x-y) \ ,
\eeq
reads, for $t=x_0-y_0$,
\beq\label{green2}
G_{{\rm ret}}(x,y;v)_{ab} 
= -i \theta(t)  \delta^{(3)} \left({\bf x}-{\bf y}-{\bf v}t\right)
   \exp (-\gamma t)\,U_{ab}(x, y) \ .
\eeq
and the iterative solution to the Boltzmann-Langevin equation 
obtains as\footnote{We thank D. B\"odeker for pointing out a mistake 
in the reasoning presented in an earlier version.}
\begin{mathletters}\label{solutionB}
\bea
\label{NA-Ohm-J}
{\cal J}^{\rho}_{(0)} (x,v) 
&=& \int^{\infty}_0 d \tau \exp (-\gamma \tau)
       U(x,x-v \tau) \left\{  -m^2_Dv^\rho v^j F_{j 0} (x-v\tau)
       + \zeta^\rho(x-v\tau,v)  \right\}
\\
{\cal J}^\rho _{(n)}(x,v) 
&=& \gamma \int^{\infty}_0 d \tau \exp (-\gamma \tau)
        U(x,x-v \tau) v^\rho
        \int\ \frac{d\Omega_{{\bf v}'}}{4\pi}\, {\cal K}(\vv,\vv') 
                                     {\cal J}^0_{(n-1)}(x-v\tau,v') \ .
\eea
\end{mathletters}%
This expansion in consistent with current conservation, if the angle 
average of ${\cal J}^0_{(n)}$ vanishes for some $n$. This follows
from taking the temporal  component of (\ref{seriesB}) and averaging 
the equation over ${\bf v}$, to find
\begin{mathletters}
\bea
D_0 J^0 _{(0)} + D_i J^i_{(0)} 
&=& - \gamma J^0 _{(0)}  \ , \\
D_0 J^0 _{(n)} + D_i J^i_{(n)} 
&=& \gamma J^0 _{(n-1)} - \gamma J^0 _{(n)}  \ ,
\eea
for the individual contributions, and
\beq
    D_\mu \left(  J^\mu_{(0)} 
                + J^\mu_{(1)}
                + \ldots 
                + J^\mu_{(n)} \right)
  + \gamma J^0 _{(n)}   = 0   \ ,
\eeq
\end{mathletters}%
for their sum, which is consistent if $\gamma J^0_{(n)}$ is vanishing 
for some $n$. 

To leading order, the ultra-soft colour current ${\cal J}^i_{(0)}(x,v)$ 
in \eq{NA-Ohm-J} has the same functional dependence on the field strength 
and on the parallel transporter as the soft colour current 
(\ref{class-HTL-curr}). There is, however, 
an additional damping factor $e^{-\gamma \tau}$ in the integrand.

We shall now consider the overdamped regime (or quasi-local limit) of 
the above equations. Consider the mean field currents \eq{solutionB}. 
The terms contributing to these currents are exponentially suppressed for 
times $\tau $ much larger than the characteristic time scale $1/\gamma$. 
On the other hand, the fields occurring in the integrand vary typically 
very slowly, that is on time scales $\ll 1/m_{\rm D}$. Thus, the 
quasi-local limit consists in the approximation
\beq
 U_{ab}(x,x-v \tau) \approx U_{ab}(x,x) = \delta_{ab} \ , 
\quad  F_{j 0}  (x-v\tau) \approx   F_{j 0}  (x) \ .
\eeq
In this case the remaining integration can be performed. The solution 
for the spatial current $J^i(x)$ stems entirely from the leading order 
term \eq{NA-Ohm-J}. All higher order corrections vanish, because 
they are proportional to
\beq
\int\frac{d\Omega_\vv}{4\pi}\,\vv \,{\cal K}(\vv,\vv')=0\ .
\eeq
Therefore, we obtain for the spatial current from \eq{solutionB} 
\begin{mathletters}
\begin{equation}
\label{Jsigma}
J^i_a         =  \sigma  E^i _a + \nu^i_a \ ,  \quad\quad
\quad\quad \sigma =  \frac{4 \pi m_D^2}{3Ng^2T \ln \left(1/g\right)}\ ,
\end{equation}
where $\sigma$ denotes the colour conductivity of the plasma. The noise 
term becomes
\beq
\label{l-noise}
\nu(x)=\01{\gamma}\int \0{d\Omega_\vv}{4\pi} \zeta(x,v)\ ,
\quad\quad\quad\quad
\left\langle \nu^i_{a}(x)\ \nu^j_{b}(y)\right\rangle 
=  2\, T \,\sigma\, \delta^{ij}\, \delta_{ab}\, \delta^{(4)}(x-y) 
\eeq
\end{mathletters}%
The noise term appearing in the Yang-Mills equation becomes white noise 
within this last approximation. The fluctuation-dissipation 
theorem is fulfilled because the strength of the noise-noise 
correlator \eq{l-noise} is precisely given by the dissipative term 
of \eq{Jsigma}. This is the simplest form of the FDT. 
The colour conductivity in the quasi-local limit has been discussed by 
several authors in the literature \cite{Arnold,Selikhov2}.

For the complete set of gauge field equations in the quasi-local limit we 
still need to know $J^0(x)$. The iterative solution \eq{solutionB} gives 
$J^0_{(n)}(x)=0$ to any finite order. Therefore, we use instead the
un-approximated dynamical equation for ${\cal J}^0(x,v)$, which yields, 
averaged 
over the directions of $\vv$, current conservation. In combination with 
the solution \eq{Jsigma} for the spatial current, the Boltzmann-Langevin 
equation \eq{fin-set} finally becomes
\begin{mathletters}\label{BEQ}
\begin{eqnarray}
D_\mu  F^{\mu i} & = & \sigma  E^i  + \nu^i \ , \\
D_i E^i & = & J^0 \\
D_0 J^{0}& =&  -\sigma J^0  - D_i \nu^i  \ .
\end{eqnarray}
\end{mathletters}%
It is worth pointing out that already in the leading logarithmic 
approximation the noise term appearing in the Yang Mills equation 
is not white, except in the local limit \eq{l-noise}. 
The noise in the Boltzmann-Langevin equation, on the other hand, is white 
(see (\ref{noi-Bolt})), when averaged over the directions of $\vv$.

For numerical computations, which can in principle take into account 
the non-localities of the problem, it might be more convenient to work 
with the  two set of equations (\ref{fin-set}), rather than with a 
non-local stochastic gauge field equation.

\section{The quantum plasma close to equilibrium}\label{quantum}

\subsection{The quantum plasma from transport theory}

Up to now we have made an entirely classical derivation of a Boltzmann
equation with collision integrals and stochastic sources, and we have 
finally derived the mean gauge field equations. The basic ingredients for 
such a derivation were the classical equations of motion and the classical 
statistical averages introduced in sect.~\ref{stati-sec}. The following 
natural step is quantizing the whole procedure in order to obtain quantum 
Boltzmann equations and the corresponding  mean gauge field equations.
 
In order to  quantize this formulation one has to abandon the concept of 
classical trajectories, and introduce commutators for all the canonical 
conjugate pairs of variables. One should also take into account that 
quantum particles  are indistinguishable. The natural formulation of the 
quantum problem is then given in terms of Wigner functions and density 
matrices. We will not present a  rigorous discussion of the quantum 
counterpart of our formulation, but rather present a minimal set of 
changes in our equations which allow to consider also quantum plasmas 
close to equilibrium. We leave for a future project a much more rigorous 
discussion based on first principles of the quantum formulation of the 
problem.

Our starting observation is that even in quantum plasmas the physics 
occurring at soft and ultra-soft scales can be encoded into classical 
or semiclassical equations. The reason for this is that the occupation 
number for soft modes close to equilibrium is very large, suggesting 
that a description in terms of classical equations  might also be valid 
to describe the physics of the soft and ultra-soft scales in the quantum 
plasmas.

Therefore, in order to consider a quantum plasma, we shall need 
to make several changes. The first step consists in expanding the mean 
distribution function around the appropriate quantum statistical 
distribution function. For a plasma close to equilibrium, these are 
given by \eq{eq-quan0}. We also change the normalisation of the 
distribution functions as indicated in sect.~\ref{scales}.

This suffices to obtain the correct quantum value for the Debye mass, 
appearing in the Vlasov equation, and thus to reproduce fully the 
HTL effective theory in the leading order in $g$ \cite{KLLM}.

When fluctuations are considered as well, it is equally important to 
modify the classical correlator \eq{average} to the corresponding 
quantum statistical one. For bosons, and for every internal degree 
of freedom, one has 
\bea
\langle\delta f_{{\bf x},p,Q} \ \delta f_{{\bf x}',p',Q'}\rangle 
&=& (2\pi)^3 \delta^{(3)}({\bf x}-{\bf x}')
             \delta^{(3)}({\bf p}-{\bf p}') 
             \delta (Q- Q') \bar f_{\rm B} (1 + \bar f_{\rm B})
\nonumber\\ &&\label{dfdf-B}
             + {\tilde g}^{\rm B}_2({\bf x},p,Q;{\bf x}',p',Q') \ ,
\eea
while the corresponding correlator for fermions is 
\bea
\langle\delta f_{{\bf x},p,Q} \ \delta f_{{\bf x}',p',Q'}\rangle 
&=& (2\pi)^3\delta^{(3)}({\bf x}-{\bf x}')
            \delta^{(3)}({\bf p}-{\bf p}') 
            \delta (Q- Q') \bar f_{\rm F} (1 - \bar f_{\rm F})
\nonumber\\ &&\label{dfdf-F}
             + {\tilde g}^{\rm F}_2({\bf x},p,Q;{\bf x}',p',Q') \ . 
\eea
The functions ${\tilde g}^{\rm B/F}_2$ are related to the 
bosonic/fermionic two-particle correlation function. The above relations 
should be derived from first principles in a similar way as our
equation (\ref{average}). In the limit $\bar f_{\rm B/F}\ll 1$ 
they reduce to the correct classical value. It has also to be pointed 
out that the correlators \eq{dfdf-B} and \eq{dfdf-F} have been derived 
for the case of both an ideal gas of bosons and ideal gas of fermions 
close to equilibrium, matching the change described above. This can be 
taken as the correct answer in the case that the non-Abelian interactions  
can be treated perturbatively.

With the above in mind, we can now describe the minimal set of changes 
to our computations of sect.~\ref{close2eq} which allows to treat the 
quark-gluon plasma close to equilibrium. We will consider  gluons in the 
adjoint representation with $C_2 = N$, and $N_F$ quarks and $N_F$ 
antiquarks in the fundamental representation, with $C_2 =1/2$.  All 
particles carry two helicities. We will neglect the masses of the 
quarks, $m \ll T$.

The value of the the quantum Debye mass squared becomes
\beq
m_D^2 = - \frac{g^2 }{\pi^2} \int^\infty_0 dp p^2  
\left( N   \frac{d \bar f^{\rm eq}_{\rm B}}{d p} 
     + N_F \frac{d \bar f^{\rm eq}_{\rm F}}{d p} \right)
\ .
\eeq
Evaluating explicitly the integral, one finds 
$m^2_D = g^2 T^2 (2N+ N_F)/6$.

The correlator of colour currents densities are then modified, according 
to the changes mentioned above. Since we now consider different species 
of particles in (\ref{cur-cor}) $C_2 B_C$ is replaced by a sum over
different species of particles. For the quark gluon plasma, all our 
equations of sect.~\ref{close2eq} remain valid if we replace $C_2 B_C$ by
\beq
\sum_{\hbox{\tiny species}} 
C_2 B_C= \frac{ 2 N}{\pi} \int^\infty_0 dp p^2 
\bar f^{\rm eq}_{\rm B} (1 + \bar f^{\rm eq}_{\rm B})
       + \frac{ 2 N_F}{\pi} \int^\infty_0 dp p^2 
\bar f^{\rm eq}_{\rm F} (1 - \bar f^{\rm eq}_{\rm F}) \ .
\eeq
It is curious that the relation (\ref{magical-rel}) for the quantum 
values of the above quantities remains unchanged, that is, 
\beq
2 g^2 \sum_{\hbox{\tiny species}} 
C_2 B_C = 4 \pi T m^2_D
\eeq
holds true also in the quark-gluon plasma. Since this combination appears
in front of all our collision integrals, we find a universal value for 
the coefficient of  (\ref{leading-log}), $\gamma$ for both the classical 
and quantum plasmas. This is always the case in the leading logarithmic 
approximation, if the IR cutoff used is of order $g m_D$, where $m_D$ 
would correspond to the classical or quantum Debye mass, respectively.
The value $\gamma/2$ can be identified with the damping rate of a hard 
transverse gluon \cite{damp-rat-P}. 

With these observations, one does not need to repeat the computations that 
we performed in sect.~\ref{close2eq}. In particular, the final mean field
equations of sect.~\ref{mean} only change in the value of the Debye mass.

\subsection{Comparison to related work}

Let us briefly comment on some related work. A similar philosophy to ours 
has already been followed by Selikhov \cite{Selikhov}. He used the
semiclassical limit of quantum transport equations for the Wigner 
functions associated to gluons and quarks, which reduce to our starting 
classical transport equations. He used a procedure of splitting both the 
Wigner functions and vector gauge fields into mean values and statistical 
fluctuations. A key point is how the statistical correlator of 
fluctuations in a quantum framework can be derived. Selikhov relied on the 
same type of statistical correlator as derived in  (\ref{cur-cor}). 
However, it should be stressed that this statistical correlator is only 
correct in the pure classical framework, for classical statistics.
This can not reproduce the correct prefactors of the quantum collision
integral. Also, the FDT is not satisfied in this case. Instead, the 
correct correlators are given by \eq{dfdf-B} and \eq{dfdf-F}. Also, the 
colour current he found is not covariantly conserved. This is so 
because the non-local term in the collision integral \eq{leading-log}, 
proportional to ${\cal K}(\vv,\vv')$, has been neglected.

The first to derive the mean field equations (\ref{fin-set}) and the 
related noise correlator \eq{noi-Bolt-0} for the quantum plasma was 
B\"odeker \cite{Bodeker}. His approach uses the local version of the 
HTL effective action as the starting point to integrate-out 
the modes with momenta about the Debye mass. He profits from the 
observation that the soft field modes behave classically. This allows 
the definition of a classical thermal average with a weight given by 
the HTL Hamiltonian. Although the starting point and the techniques 
involved appear to be quite different from our approach, it remains 
intriguing that the integrating-out of momentum modes within a field
theory results in the same effective theory as integrating-out 
statistical fluctuations within the present transport theory approach.

A different line has been followed by Arnold, Son and Yaffe 
\cite{Arnold}, who realized that B\"odeker's effective theory has a 
physical interpretation in terms of kinetic equations. They managed
to derive the relevant collision term of the Boltzmann equation on 
phenomenological grounds. 

Very recently, the quantum collision integral has been obtained  
within a quantum field theoretical setting  by Blaizot and Iancu 
\cite{BI-preprint}. The derivation relies on a gauge covariant 
derivative expansion of the quantum field equations. While conceptually 
very different to our approach \cite{LM}, the approximations used in 
\cite{BI-preprint} are very similar to the ones we performed. This is 
maybe not too surprising after all, as the explicit computations in
both approaches are based on a consistent expansion in powers of the 
gauge coupling constant. It seems only that the concept of statistical 
fluctuations has not been introduced in \cite{BI-preprint}, which may 
be a reason for why the source of stochastic noise, necessary for a 
correct macroscopic description of the plasma, has not yet been 
identified. The stochastic noise can probably be derived by considering 
the effects of higher order correlation functions \cite{Iancu}.

Based on our approach \cite{LM}, the quantum collision integral for 
the transport equation has been obtained as well by Valle \cite{Valle}. 
He started from the HTL effective theory and found the correct 
damping rate after imposing a fluctuation-dissipation relation. 
The noise term in his final equations is however missing, which again 
would entail that the system abandons equilibrium.  

Finally, it is interesting that the collision integral can be 
interpreted in terms of Feynman diagrams \cite{Bodeker,BI-preprint}. 
B\"odeker also made a diagrammatic derivation  of his effective 
theory \cite{Bodeker}. This is a much more lengthy and cumbersome task,
and shows, on the other hand, the very efficiency of a kinetic approach,
as it corresponds to a re-organisation of the perturbative series.

\section{Discussion}\label{dis}

We have presented a self-consistent approach to study classical 
non-Abelian plasmas. Let us summarize here again our starting assumptions 
to derive the effective transport equations. A system of point particles
carrying non-Abelian charges is considered. Their microscopic equations of
motion are the Wong equations. In order to describe an ensemble of these 
particles, we introduced an ensemble average, which takes also the colour 
charges as dynamical variables into account. This yields finally a set
of transport equations for both mean quantities and statistical 
fluctuations, and gives a recipe to obtain explicitly the collision 
integrals for macroscopic transport equations. This approach is consistent 
with gauge invariance, and admits systematic approximations. Most 
particularly, it does not rely on close-to-equilibrium situations. These 
techniques, applied since long to Abelian plasmas, have never been fully 
exploited for the non-Abelian case. Our approach is aimed at closing this
gap in the literature of non-Abelian plasmas.

We applied this method to non-Abelian plasmas close to thermal 
equilibrium. A sufficiently small gauge coupling parameter is at the 
basis for a systematic expansion of the dynamical equations. Neglecting 
fluctuations yields to leading order the known non-local expression for 
the soft current in terms of the soft gauge fields (HTL approximation).
Integrating-out, in addition, the fluctuations to leading logarithmic 
order, that is with momenta about $m_D$, results in a Boltzmann-Langevin 
equation for the ultra-soft modes, 
\[
v^\mu  D_\mu {\cal J}^\rho(x,v)
=  -  m^2_D v^\rho v^\mu  F_{\mu0} (x)
   - \gamma\  v^\rho
      \int \frac{d\Omega_{{\bf v}'}}{4\pi}\, 
           {\cal I}(\vv,\vv') {\cal J}^0(x,v') 
   + \zeta^{\rho}(x,v)\, .
\]
It contains a collision term and a related noise term,  with 
$\gamma=g^2 N T \ln{(1/g)}/4 \pi$, while the stochastic source 
$\zeta$ obeys
\[
\llangle \zeta^\mu_a (x,v)\, 
         \zeta^\nu_b (y,v') \rrangle 
= 2\,\gamma\,T\,m^2_D 
   \,v^\mu v'^\nu\, {\cal I}(\vv,\vv') \,
   \delta_{ab}\, \delta^{(4)} (x-y) \ .
\]
The above is completed with the Yang-Mills equations.
Solving the Boltzmann-Langevin equation (see sect.~\ref{mean}), one 
obtains the Yang-Mills equation for the ultra-soft fields 
\[
[D_\mu, F^{\mu\nu}](x)
 = \int\!\frac{d\Omega_{{\bf v}}}{4\pi}\,{\cal J}^\nu(x,v)\ . 
\]
Surprisingly, the dynamical equations are the same for both classical and
quantum plasmas, the only difference being the value for the
Debye mass. This conclusion relies also on the use of an infrared cut-off
of order $g m_D$, where $m_D$ is the classical or quantum Debye mass.
For the quantum plasmas, our result agrees with the quantum collision 
integrals found in the literature using different methods 
\cite{Bodeker,Arnold,BI-preprint}. The main effect of the fluctuations 
with momenta about the Debye mass is the introduction of a damping term 
and a source of stochastic noise into the above expression. Note also 
that the damping coefficient $\gamma$ is the same for classical and 
quantum plasmas.

Our work establishes a link even beyond the one-loop level between
the classical transport theory approach as presented here, and a full
quantum field theoretical treatment. It would be most desirable if this
connection could be further substantiated. This should also yield
a quantitative criterion for the applicability of the 
-technically speaking- much simpler approach based on the 
classical point particle picture. 

Let us finally emphasize that the same IR problems, which are due to
the unscreened magnetic modes, appear both for classical and quantum
plasmas. This suggests that the solution for these IR divergences might 
also be the same in the two cases. Therefore, it seems profitable to seek 
for a solution to this problem in the much simpler framework of classical 
transport theory, rather than in a quantum field theoretical approach.

\acknowledgements
We wish to thank Dietrich B\"odeker for discussions.

\appendix

\section{Darboux variables}\label{Measure}

The statistical averages defined in sect.~\ref{stati-sec} have to be
performed in phase space. The colour charges $Q_a$ are not real 
phase space variables \cite{KLLM}.  It is possible to define 
the set of Darboux variables associated to the $Q_a$ charges.

For $SU(2)$ we define the new set of variables $\phi, \pi, J$ by the 
transformation \cite{KLLM}
\beq
Q_1= \cos{\phi} \sqrt{J^2 - \pi^2} \, \qquad 
Q_2= \sin{\phi} \sqrt{J^2 - \pi^2} \, \qquad Q_3 = \pi
\eeq
where $\pi$ is bounded by $-J \leq \pi \leq J$. The variables $\phi, \pi$ 
form a canonically conjugate pair, while $J$ is fixed by the value of the
quadratic Casimir, which is constant under the dynamical evolution.
 One can define Poisson brackets with these canonical variables,
under which the colour charges form a representation of $SU(2)$, 
$\{Q_a,Q_b\}_{PB} = f_{abc} Q_c$.
With the above change of variables, one can easily fix the value of the 
representation normalization constant $c_R$ introduced in (\ref{col-mes}).
From the condition $\int dQ =1$ one finds $c_R = 1/{2 \pi \sqrt{q_2}}$.
From the condition $\int dQ Q_a Q_b = C_2 \delta_{ab}$ one gets 
$q_2 = 3 C_2$. This entirely fixes the value of $c_R$ as a function 
of $C_2$.

The Darboux variables associated to $SU(3)$ were defined in \cite{KLLM},
and will not be discussed explicitly here.

We should also comment that in the pure classical framework, $C_2$ 
carries dimensions of $\hbar c$.  After quantization, the quadratic 
Casimirs should take quantized values proportional to $\hbar$. The 
Poisson brackets then have to be replaced by commutators.

\section{Consistent current conservation}\label{identity}

In this appendix we verify explicitly the identity
\beq   \label{general-c}
 0=   [\bar D_\mu,J_{{\mbox{\tiny fluc}}}^\mu ]
   +g [a_\mu,\de J^\mu]
   +g [a_\mu,\langle J_{{\mbox{\tiny fluc}}}^\mu \rangle] \ ,
\eeq
which is at the basis for the proof of the consistent current 
conservation of both the mean field and the fluctuation current in 
sect.~\ref{consistentcurrent}. The following check is algebraic, and it 
will  make use of symmetry arguments like the antisymmetry of the 
commutator and the tensors $\bar F_{\mu\nu}, f_{\mu\nu}$, and of the 
cyclic identity $[t_a,[t_b,t_c]]+[t_b,[t_c,t_a]]+[t_c,[t_a,t_b]]=0$. The 
identity $[\bar D_\mu,\bar D_\nu]=g\bar F_{\mu\nu}$ is employed as well. 
To simplify the computation, we will seperate the fluctuation part of 
the field strength \eq{NA-deltaF-c} into the term linear and quadratic 
in $a$, according to
\beq
f_{\mu\nu}   = f_{1,\mu\nu}+f_{2,\mu\nu}               \ ,\qquad 
f_{1,\mu\nu} =   [\bar D_\mu,a_\nu]-[\bar D_\nu,a_\mu] \ ,\qquad 
f_{2,\mu\nu} = g [a_\mu,a_\nu]\ .
\eeq
Recall furthermore, using \eq{M-NA-Jfluc} and \eq{M-NAJ-2}, that
\bea
J_{{\mbox{\tiny fluc}}}^\mu&=&
               [\bar D_\nu,f^{\nu\mu}_2] 
             +g[a_\nu,f^{\nu\mu}_1+f^{\nu\mu}_2]          \label{Jfluc}
\\ 
\de J^\mu &=& [\bar D_\nu,f^{\nu\mu}_1]
             + g[a_\nu,\bar F^{\nu\mu}]
             +J_{{\mbox{\tiny fluc}}}^\mu
             -\llangle J_{{\mbox{\tiny fluc}}}^\mu\rrangle \label{deJ}
\eea
are functions of the fluctuation field $a$. The first term of 
\eq{general-c} reads, after inserting  $J_{{\mbox{\tiny fluc}}}$ 
from \eq{Jfluc}, 
\bea
\label{lhs}
[\bar D_\mu,J_{{\mbox{\tiny fluc}}}^\mu]&=& 
               [\bar D_\nu,[\bar D_\mu,f^{\mu\nu}_{2}]]
             +g[\bar D_\nu,[a_\mu,f^{\mu\nu}_{1}]]
             +g[\bar D_\nu,[a_\mu,f^{\mu\nu}_{2}]]\ .
\eea
Using $\de J$ from \eq{deJ}, it follows for the second term of 
\eq{general-c}
\bea
\label{rhs}
g[a_\mu,\de J^\mu]&=& g^2[a_\nu,[a_\mu,\bar F^{\mu\nu}]]
        +g[a_\nu,[\bar D_\mu,f^{\mu\nu}_{1}]]
        +g\left[a_\nu,J_{\mbox{\tiny fluc}}^{\nu}\right]
        -g\left[a_\nu,\langle J_{\mbox{\tiny fluc}}^{\nu}\rangle\right]
\eea
The last term of \eq{rhs} will be canceled by the last term in 
\eq{general-c}. We show now that the first three terms of \eq{lhs} 
and \eq{rhs} do cancel one by one. The first term in \eq{lhs} can be 
re-written as
\beq
[\bar D_\nu,[\bar D_\mu,f^{\mu\nu}_{2}]] 
              =  [[\bar D_\nu,\bar D_\mu],f^{\mu\nu}_2] 
                -[\bar D_\nu,[\bar D_\mu,f^{\mu\nu}_2]]  
              =  \s012 g [\bar F_{\nu\mu},f^{\mu\nu}_2]
\eeq
Similarly, the first term of \eq{rhs} yields
\beq
g^2[a_\nu,[a_\mu,\bar F^{\mu\nu}]]  
              = -g^2[\bar F^{\mu\nu},[a_\nu,a_\mu]]
                -g^2[a_\nu,[a_\mu,\bar F^{\mu\nu}]] 
              = -\s012 g [\bar F_{\mu\nu},f^{\nu\mu}_2]\ .
\eeq
For the second term in \eq{lhs} we have
\beq
g[\bar D_\nu,[a_\mu,f^{\mu\nu}_{1}]]  
              =  g[a_\mu,[\bar D_\nu,f^{\mu\nu}_{1}]]
                +g[[\bar D_\nu,a_\mu],f^{\mu\nu}_{1}] 
              = -g[a_\mu,[\bar D_\nu,f^{\nu\mu}_{1}]] \ ,
\eeq
which equals (minus) the second term of \eq{rhs}. Finally, consider the
third term of \eq{rhs},
\bea
g\left[a_\nu,J_{\mbox{\tiny fluc}}^{\nu}\right]
&=&    g^2[a_\nu,[\bar D_\mu,[a^\mu,a^\nu]]] 
     + g^2[a_\nu,[a_\mu,f^{\mu\nu}]]           \nonumber\\
&=&    \s012 g[f^{\mu\nu}_2,f_{1,\mu\nu}]
     - g[\bar D_\mu,[f_2^{\mu\nu},a_\nu]]
     -\s012 g[f^{\mu\nu}_2,f_{1,\mu\nu}]       \nonumber\\
&=&  -g[\bar D_\mu,[a_\nu,f_2^{\nu\mu}]]
\eea
which equals (minus) the third term of \eq{lhs}. This establishes
\eq{general-c}.

\newpage

\end{document}